\documentclass[sigconf]{acmart}
\usepackage{float}
\usepackage{tabularx}
\AtBeginDocument{%
  \providecommand\BibTeX{{%
    \normalfont B\kern-0.5em{\scshape i\kern-0.25em b}\kern-0.8em\TeX}}}

\copyrightyear{2023}
\acmYear{2023}
\setcopyright{rightsretained}
\acmConference[UIST '23]{The 36th Annual ACM Symposium on User Interface Software and Technology}{October 29-November 1, 2023}{San Francisco, CA, USA}
\acmBooktitle{The 36th Annual ACM Symposium on User Interface Software and Technology (UIST '23), October 29-November 1, 2023, San Francisco, CA, USA}
\acmDOI{10.1145/3586183.3606821}
\acmISBN{979-8-4007-0132-0/23/10}

\usepackage{soul}
\usepackage{hyperref}

\renewcommand\hl[1]{#1}

\hypersetup{
pdftitle={Template},
pdfsubject={Human-Computer Interaction},
pdfauthor={},
pdfkeywords={}
}

\AtBeginDocument{%
  \providecommand\BibTeX{{%
    Bib\TeX}}}

\begin{document}
\settopmatter{printfolios=true}

\newcommand{\system}{SmartPoser}
\title{\system: Arm Pose Estimation with a Smartphone and Smartwatch Using UWB and IMU Data}


\author{Nathan DeVrio}
\authornote{Both authors contributed equally.}
\affiliation{
 \institution{Carnegie Mellon University}
 \city{Pittsburgh}
 \state{PA}
 \country{USA}}
\email{ndevrio@cmu.edu}

\author{Vimal Mollyn}
\authornotemark[1]
\affiliation{
 \institution{Carnegie Mellon University}
 \city{Pittsburgh}
 \state{PA}
 \country{USA}}
\email{vmollyn@andrew.cmu.edu}

\author{Chris Harrison}
\affiliation{
 \institution{Carnegie Mellon University}
 \city{Pittsburgh}
 \state{PA}
 \country{USA}}
\email{chris.harrison@cs.cmu.edu}

\renewcommand{\shortauthors}{DeVrio \& Mollyn, et al.}

\begin{abstract}
The ability to track a user's arm pose could be valuable in a wide range of applications, including fitness, rehabilitation, augmented reality input, life logging, and context-aware assistants. Unfortunately, this capability is not readily available to consumers. Systems either require cameras, which carry privacy issues, or utilize multiple worn IMUs or markers. In this work, we describe how an off-the-shelf smartphone and smartwatch can work together to accurately estimate arm pose. Moving beyond prior work, we take advantage of more recent ultra-wideband (UWB) functionality on these devices to capture absolute distance between the two devices. This measurement is the perfect complement to inertial data, which is relative and suffers from drift. We quantify the performance of our software-only approach using off-the-shelf devices, showing it can estimate the wrist and elbow joints with a \hl{median positional error of 11.0~cm}, without the user having to provide training data. 
\end{abstract}

\ccsdesc[500]{Human-centered computing~Mobile devices}

\keywords{Smartwatch, sensing, hand gestures, body pose, mobile devices, interaction techniques. }

\begin{teaserfigure}
\centering
\resizebox{1.0\textwidth}{!}{
    \includegraphics[width=\textwidth]{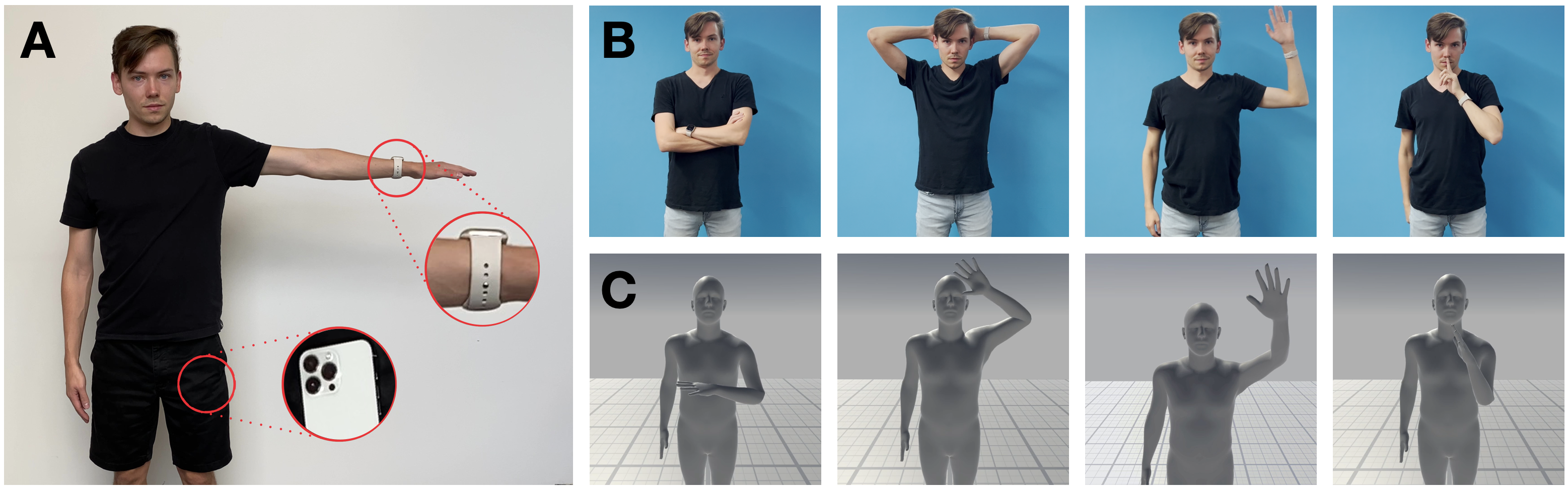}
}
\caption{SmartPoser fuses UWB distance and IMU data from an off-the-shelf smartphone and smartwatch (A) to track a wearer's real-time arm pose (B \& C).}
\Description[A panel of photos showing a user with a smartphone and smartwatch and example arm poses that SmartPoser can capture.]{A series of images showing our SmartPoser system. On the left is a large picture of the hardware in our system, a watch on the user's wrist and a phone in their pocket. On the right are two rows of images. In the top row a user is seen performing various arm actions, and on the bottom row matching visualizations of predicted arm pose are shown.}
\label{fig:teaser}
\end{teaserfigure}
\maketitle

\section{Introduction}
Despite being worn on the wrist, contemporary smartwatches know more about their users' legs, lungs, and heart than their arms. For instance, smartwatches can predict locomotion mode, track total number of steps taken, and measure respiration and heart rate. Yet, beyond some very specific use cases (e.g., hand washing detection \cite{handwashingApple}, the activities of the arms and hands are unknown, despite the fact that they are the chief appendage we use to manipulate the world around us. If we could better track the arms, we could power more sophisticated applications spanning fitness \cite{wilk_multimodal_2021, ding_platform_2017}, rehabilitation \cite{baldominos_approach_2015}, life logging \cite{meyer_making_2015}, occupational training \cite{bosche_towards_2016}, and context-aware assistants \cite{rajanna_step_2014}, to name just a few domains.

Researchers have long recognized the importance of tracking arm pose to support such use cases. As we will discuss in the Related Work section, a wide variety of methods have been brought to bear on this problem. Much prior work has utilized external, fixed equipment (most often cameras), but this immediately precludes mobile use. Next most common are worn sensor arrays or suits, which are comparatively more practical, but still face significant consumer adoption headwinds. Ideally, we could track a user's arm pose with devices they already own and routinely carry with them \cite{ichikawa_wheres_2005,redmayne_wheres_2017}, relying on no external infrastructure. 

In this work, we present SmartPoser, a new arm pose tracking method that uses only an off-the-shelf smartwatch and smartphone (Figure \ref{fig:teaser}). In other words, our technique is software only, and could be enabled on recent devices as an over-the-air update or app download. SmartPoser also allows users to be on the move while it performs its tracking, a capability not found in many prior approaches \cite{shen_i_2016, zhou_limbmotion}. Taken together, these advances could significantly improve the ease-of-use and practicality of mobile arm-tracking applications. 

Uniquely, we take advantage of recent ultra-wideband (UWB) functionality to capture the distance between the smartwatch and smartphone. We fuse this with more conventional IMU-reported data, specifically watch and phone orientation and acceleration. Importantly, the UWB measurements are absolute, and less drift-prone as compared to relative inertial data used in most prior work. As a consequence, our tracking output is more stable and accurate. In our user study, our system achieved a \hl{median positional error of 11.0~cm for the wrist and elbow joints}, which is comparable to prior work requiring uncommon hardware. 

The contributions of this work are as follows: 1) A self-contained and software-only approach for arm pose tracking combining UWB and IMU data, which is unique in the literature. 2) A functional, real-time implementation using two common off-the-shelf devices. 3) User studies that show our system achieves the best tracking accuracy among systems that do not require special user instrumentation or external infrastructure. 4) Open-sourced, synchronized UWB/IMU/Kinect data, processing pipeline, and trained models. https://github.com/FIGLAB/SmartPoser.

\section{Related Work}
In this section, we review past systems for arm pose tracking. In addition to arm tracking, we also discuss systems for whole-body pose tracking that have used comparable sensing technologies. We first cover systems that use UWB for pose tracking, before moving to IMU-based systems. We then conclude with systems most similar to our own --- ones that combine inertial data with some form of distance-based measurement.

\subsection{UWB-Based Pose Tracking}
Our system takes advantage of ultra-wideband (UWB) ranging. This technique uses the time-of-flight of a challenge-response transmission to determine the distance between two devices. Unlike other more popular RF technologies, such as WiFi or Bluetooth, UWB operates at a very low energy level and instead of varying frequency or power, the technology transmits information at specific times using a large bandwidth. This results in multiple beneficial properties such as high immunity to multipath fading, high data throughput, and excellent time-domain resolution, allowing for more accurate localization and tracking, down to centimeters \cite{grobwindhager_uwb}.

Due to the strengths just described, UWB has become a popular approach for pose tracking systems due to its ability to accurately measure the distance between devices in space. A common setup is to place UWB beacons on the limbs and torso of a user and instrument an environment with one or more receivers to measure the time-of-flight of signals transmitted from the beacons \cite{yang_environment_2022, mekonnen_constrained_2010}. Researchers have also created a UWB radar with a large, MIMO antenna array that can capture a radio "image" of the user's pose \cite{song_through-wall_2021}. When compared to cameras, UWB can be more privacy-sensitive and can operate through walls and other occlusions \cite{song_through-wall_2021}. 

Beyond UWB, there have also been other RF technologies used for pose tracking, many of which share similar advantages to those just described for UWB. In comparable setups, researchers have placed a transmitter in the environment and tracked the pose of a user (often uninstrumented) by measuring signal reflections. Examples of such systems include ones that use RFID scanners and tags \cite{jin_towards_2018, wang_rf-kinect_2018}, mmWave radar \cite{sengupta_mm-pose_2020}, and even WiFi \cite{ren_gopose_2022, ren_winect_2022}. One drawback of systems that use purely RF ranging is that although they are able to measure the distance to different parts of the body, it is difficult to distinguish fine limb orientations such as the rotation of the wrist. For this reason, RF systems are often combined with IMU sensors that can provide information on limb orientation \cite{corrales_hybrid_2008, hol_tightly_2009, zihajehzadeh_uwb-aided_2015}.

\subsection{IMU-Based Wearable Pose Tracking}


The idea of tracking pose with a wearable system of inertial sensors did not surface until the miniaturization of magnetic, angular-rate, and gravity (MARG) sensors in the late 1990s \cite{bachmann_inertial_2001}. Pioneering systems had to deal with the problem of fusing measurements from multiple instruments to track limbs, all the while minimizing drift. In addition, these systems also had difficulty detecting and correcting for erroneous measurements such as magnetic disturbances \cite{bachmann_inertial_2001, marins_extended_2001}. The two dominant approaches that were developed to solve these problems include Complimentary filters by Bachman et al. \cite{bachmann_inertial_2001} and Kalman filters by Marins et al. \cite{marins_extended_2001}. While these two seminal works would develop the theoretical foundations for tracking pose with inertial data, many others went on to take these ideas and develop them into real-world systems \cite{young_use_2010, miezal_generic_2013, lisini_baldi_upper_2020, zhou_use_2008}. The commercial Xsens system can be seen as a current peak of this classical filtering and optimization-based approach \cite{roetenberg_xsens_2013}. By using a suit embedded with 17 IMUs located all over the body, this system is able to track a wide range of motions at 120 FPS \cite{roetenberg_xsens_2013}.

In recent years, a new class of systems has emerged which are able to use significantly fewer IMUs to track pose with a similar level of accuracy \cite{wei_real_time, shen_muse, yi_transpose_2021, yi_physical_2022, jiang_avatarposer_2022, butt_magnetometer_2021, shen_i_2016}. A breakthrough and common theme in these systems has been modeling inertial data over multiple frames in time, often with neural networks. A seminal system is Sparse Inertial Poser (SIP) by Marcard et al., which was able to drop the sensor burden from 17 IMUs down to 6 by using an optimization-based method \cite{von_marcard_sparse_2017}. Building upon this, Huang et al.'s Deep Inertial Poser (DIP) \cite{huang_deep_2018} removed the constraint in SIP of needing to access the entire sequence of data and instead only used a subset of nearby frames in a bidirectional RNN (Bi-RNN). More recently, TransPose \cite{yi_transpose_2021} and Physical Inertial Poser (PIP) \cite{yi_physical_2022} advanced DIP's Bi-RNN approach, improving fidelity and framerate (to 90 FPS). This was primarily achieved by decomposing the model into sub-tasks that are easier to solve and combining the outputs to get user pose and global translation. More recently, IMUPoser~\cite{mollyn_imuposer_2023} used IMUs already found in smartphones, smartwatches and earbuds to produce a best-guess pose from one to three points of instrumentation. 

\hl{Of the IMU-based systems, the most similar to ours is ArmTrak, developed by Shen et al.} \cite{shen_i_2016}. \hl{Similar to SmartPoser, the system focused on tracking arm pose (instead of full-body pose) and did this by using a single IMU on the wrist instead of multiple IMUs placed across the body. Unlike the later deep learning methods, they used a hidden Markov model to probabilistically estimate next states for arm joint positions. This approach produced good arm pose estimates using a single IMU. However, the method was slow (with reduced accuracy when run in real-time) and could not track users on the move (only evaluated with feet planted after calibration). We detail our improvements over ArmTrak, particularly tracking results, in subsection} \ref{comp_prior}.

\subsection{Fusing Inertial and Distance Measurements for Wearable Pose Tracking}

One of the greatest challenges in working with IMUs, particularly consumer-grade components, is noise and drift. For this reason, it is often advantageous to have a secondary sensor stream that can provide absolute position or orientation data. In the context of body tracking, worn systems have combined IMU data with magnetic field sensors \cite{roetenberg_ambulatory_2007, wittmann_magnetometer-based_2019}, ultrasonic transmitters and microphones \cite{liu_realtime_2011, jahren_towards_2021, vlasic_practical_2007}, infrared distance sensors \cite{xiao_wearable_2018}, micro-flow sensors \cite{liu_wearable_2020}, \hl{acoustic ranging} \cite{zhou_limbmotion}, and optical linear encoders \cite{nguyen_wearable_2011}. As previously mentioned, UWB beacons have also been used for this purpose, albeit not in a wearable form factor \cite{corrales_hybrid_2008, hol_tightly_2009, zihajehzadeh_uwb-aided_2015}. For each of these sensing approaches, there are important trade-offs to consider such as cost, robustness to occlusion, and across-environment reliability. Yet despite their differences, these sensors all share the same operating principle of capturing absolute measurements to augment the relative measurements of worn IMUs. Most importantly, all of the aforementioned systems in this section require new and special devices to be worn by the user. A key contribution of this work is showing this ranging+IMU fusion approach can now be achieved on popular commodity devices that users already own.

\section{Implementation} \label{implementation}
\begin{figure}[b]
\begin{center}
 \includegraphics[width=\linewidth]{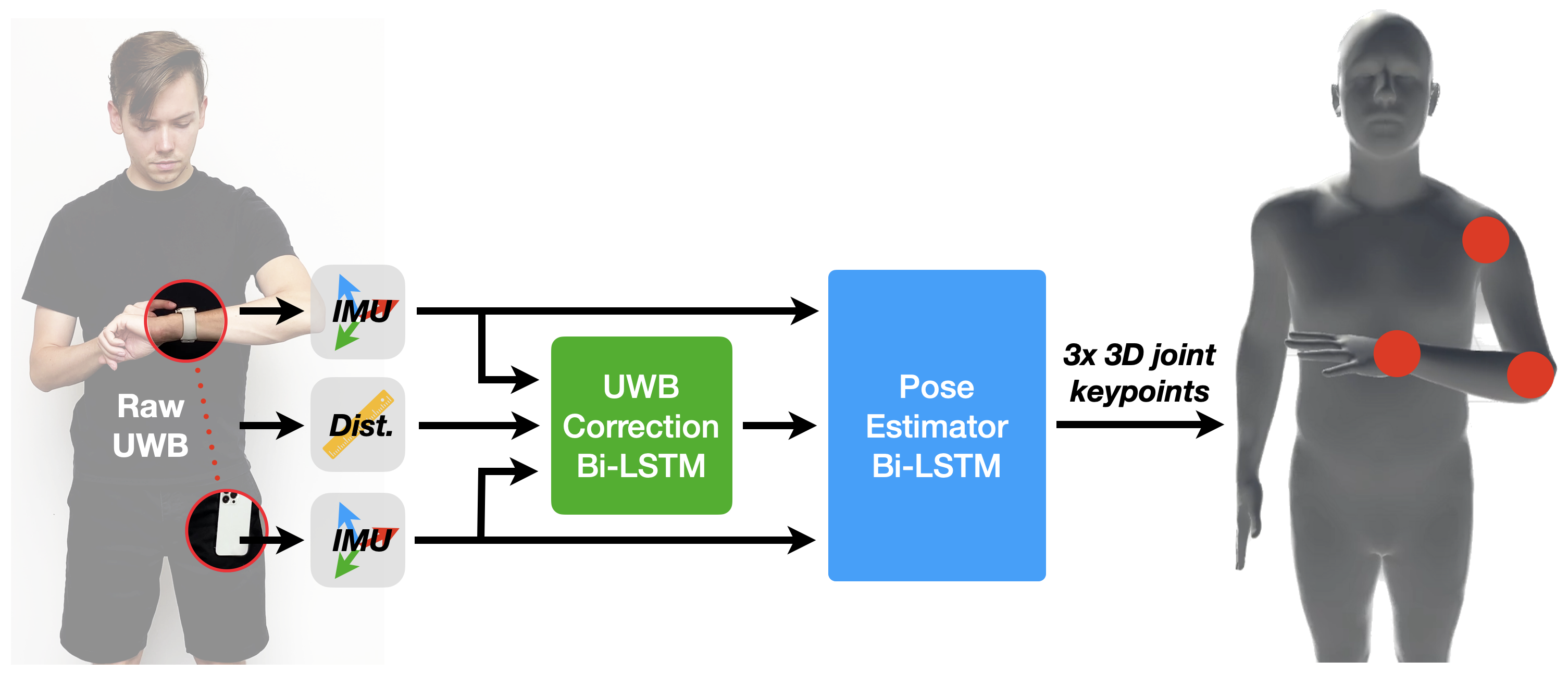}
\end{center}
  \caption{SmartPoser system overview. \hl{Our model has 46.4k parameters in total, with 3k parameters for the UWB Correction model and 43.4k parameters for the Pose Estimator model. }}
\label{fig:system_diagram}
\Description[A system diagram showing our data pipeline.]{A system diagram of the SmartPoser data pipeline. First, the relative IMU orientation between the watch and phone is calculated. This is done by subtracted phone orientation from the watch orientation. This relative orientation is then ingested along with UWB distance data into a Bi-LSTM model that outputs arm joint rotation predictions.These rotation predictions are then used to visualize the predicted pose, and passed through a forward kinematics model to get predicted joint positions. The difference between the predicted joint positions and the ground truth joint positions from a Azure Kinect are then used to calculate the loss of the system.}
\end{figure}

Our motivation in developing SmartPoser was to track arm pose using only a smartphone and smartwatch (Figure \ref{fig:teaser}). Figure \ref{fig:system_diagram} provides a high-level system overview and data flow. We now describe the core components of our system.

\subsection{Proof-of-Concept Devices}
We selected an iPhone 13 Pro (iOS 16.4) and an Apple Watch Series 7 (watchOS 9.4) as a pair of example, off-the-shelf, and popular smart devices. This choice was also motivated by Apple's mature UWB and inter-device communication APIs that facilitated our development. Android lags behind in this regard, with Google adding UWB support via a Jetpack library only in late 2022 \cite{google_uwb}. This is to support applications like lost object tracking, unlocking cars, and AR localization. Fortunately, the number of UWB-capable smart devices has steadily grown in recent years \cite{listofUWBWiki}, and the feature looks to become increasingly pervasive. 

\subsection{UWB Ranging \& IMU Data}
In our custom app, we use the iOS Nearby Interaction API \cite{noauthor_nearby_2022} to set up a UWB ranging session between the watch and phone. Once the session has been established, we receive distance updates at approximately 5 Hz. We note that iPhones offer higher sampling rates and even azimuth and elevation readings when a UWB device is located within a narrow cone-shaped field-of-view projecting from the rear of the device. Unfortunately, the field of view is too narrow for us to make practical use of this capability. 

To get data from the IMUs onboard the phone and watch, we used Apple's Core Motion API \cite{noauthor_core_2022}, which is standard across all iOS devices. Using the API, we acquire data about the current orientation quaternions and acceleration measurements of each device at roughly 25 FPS. \hl{We use this as the native frame rate of our system, upsampling the 5 Hz UWB data to the more responsive 25 Hz IMU rate by duplicating the most recent UWB value.} Finally, we note that although the two devices can estimate the gravity vector and magnetic north, we found it was more reliable to explicitly align the two devices' frames of reference. For this, we require the devices to be placed side by side, which could happen e.g., in a charging dock overnight. 

\subsection{UWB Correction}\label{sec:uwb_acc}

\begin{figure*}[t]
    \centering
    \includegraphics[width=\linewidth]{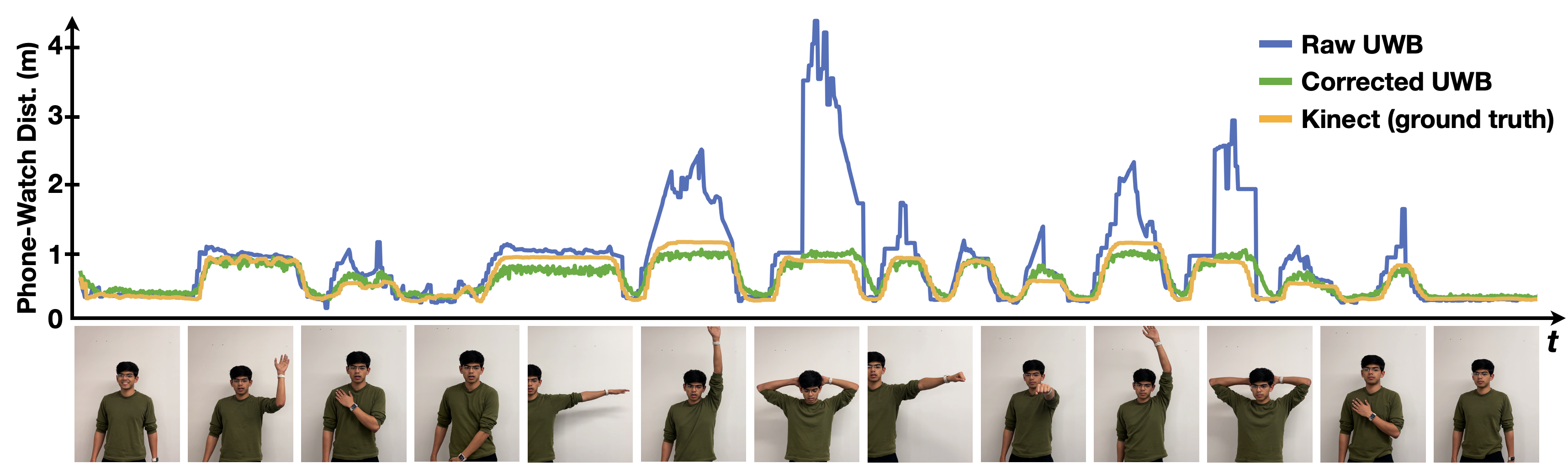}
    \caption{Example user motion sequence over one minute. Raw UWB distance is plotted in blue --- note the significant errors, far exceeding possible arm lengths, when the watch loses line-of-sight to the phone. Our corrected UWB stream is plotted in green. Ground truth distance is provided by an external Azure Kinect is shown in orange.}
    \label{fig:uwb_corrected}
    \Description[Plot showing effect of UWB-corrected distance while performing example motions.]{Plot of the phone watch distance over one minute for raw UWB, corrected UWB, and the Kinect ground truth while performing a series of user motions. At times the raw UWB separates significantly from the Kinect data, however over the course of the entire minute, the correct UWB distance tracks the Kinect value well.}
\end{figure*}

In early piloting, we saw that UWB measurements exhibited systematic offsets and characteristic noise, especially for certain arm pose regions. This is chiefly due to loss of line-of-sight between the two devices due to body occlusion (e.g., putting the arms behind the head). The example data in Figure \ref{fig:uwb_corrected} illustrates this effect. To capture this data, the authors placed the iPhone 13 Pro smartphone in their left-front pants pocket, and wore the Apple Watch Series 7 smartwatch on their left wrist. The authors then performed a range of exemplary arm motions. During this process, the phone-to-watch distance was recorded (Figure \ref{fig:uwb_corrected}, blue line). To provide a ground truth distance, we used an Azure Kinect and its native SDK to capture a 3D skeleton of the user in real-world units. Using this skeleton, we ''attached'' a virtual phone and watch, and compute the distance between the two (Figure  \ref{fig:uwb_corrected}, orange line). 

Given the offset and noise is semi-structured, we can compensate for some error, especially if we take advantage of orientation data from the IMUs, which already provide some cues as to the orientation and location of the arms. For this, we designed a UWB correction network, which we implemented as a 2-layer bi-directional LSTM with 8 hidden units each. The input to the RNN is a 125-long sequence (5 seconds of data at our system's native framerate of 25 FPS) of raw UWB distance measurements, plus phone and watch orientations (3x3 rotation matrix format) and accelerations (X/Y/Z), for a total feature dimension of 25. We scale the UWB distance measurements by the wearer's arm span. The scaling ensures that the model is able to generalize to various body sizes. We also scale the acceleration values by 30 in order to make them suitable for training, following prior work \cite{yi_transpose_2021, yi_physical_2022}. The RNN outputs a sequence of corrected distance measurements. The correcting effect of this RNN is illustrated in Figure \ref{fig:uwb_corrected}, green line. In this example sequence, mean UWB tracking error is reduced from 34.5 cm to 6.7 cm, and is considerably more stable as well.

\subsection{Pose Estimation}
Our pose estimation RNN ingests corrected UWB data along with raw IMU data (watch and phone orientation and acceleration). Taking inspiration from Deep Inertial Poser \cite{huang_deep_2018} and TransPose \cite{yi_transpose_2021}, this network is also implemented as a 2-layer bi-directional LSTM, with 32 hidden units each. Watch orientation and acceleration are normalized such that they are relative to the phone's frame of reference. \hl{Along with having an absolute distance measurement, using relative orientations helps to make the system more robust against IMU drift.} As with UWB correction, we scale the acceleration values by 30. We use the same input vector format as our UWB correction network -- a 125-length sequence of UWB distances, phone and watch orientations and accelerations. \hl{This 125-length sequence is a rolling input buffer, and new predictions are made in real-time whenever a new sample arrives in the buffer (at 25 Hz).}

This RNN outputs a 125-long arm pose sequence, containing 3D joint locations for the wrist, elbow, and shoulder. \hl{The predicted joint locations are 9 values normalized to the user's arm span, which we can "reproject" into real-world units by scaling them by the user's arm span.} Rather than taking the most recent pose prediction (i.e., the 125th frame), we use the 120th frame. Functionally, this means we are predicting 200 ms behind real-time. \hl{This approach to use the n-5th output frame was directly inspired by TransPose} \cite{yi_transpose_2021}, and offers a compromise between latency vs. accuracy. Note, during testing we also used the n-5th frame of ground truth to match with our prediction frame.

\begin{figure}[b]
\begin{center}
  \includegraphics[width=1.0\linewidth]{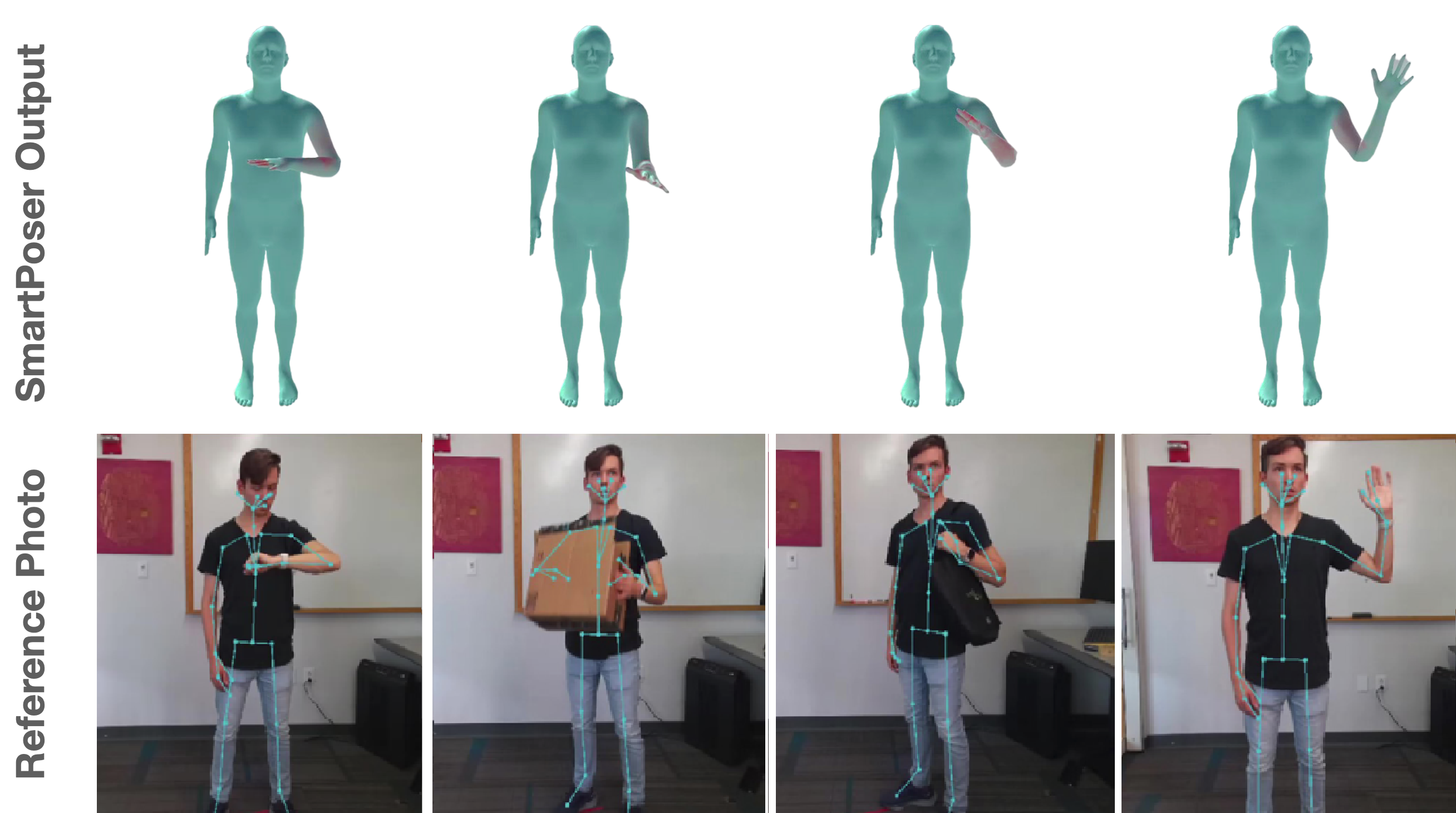}
\end{center}
 \caption{Example set of predicted poses. Vertex error (vs. Kinect ground truth) color coded. }
\label{fig:good_arm_poses}
\Description[Example photos of real poses and arm pose predictions with error colorized.]{Matched sets of rendered pose predictions and images of the reference pose overlaid with Kinect joint data. The four poses shown are checking time on watch, raising hand, carrying a box, and wearing a tote bag.}
\end{figure}

\begin{figure*}[t]
\begin{center}
 \includegraphics[width=1.0\textwidth]{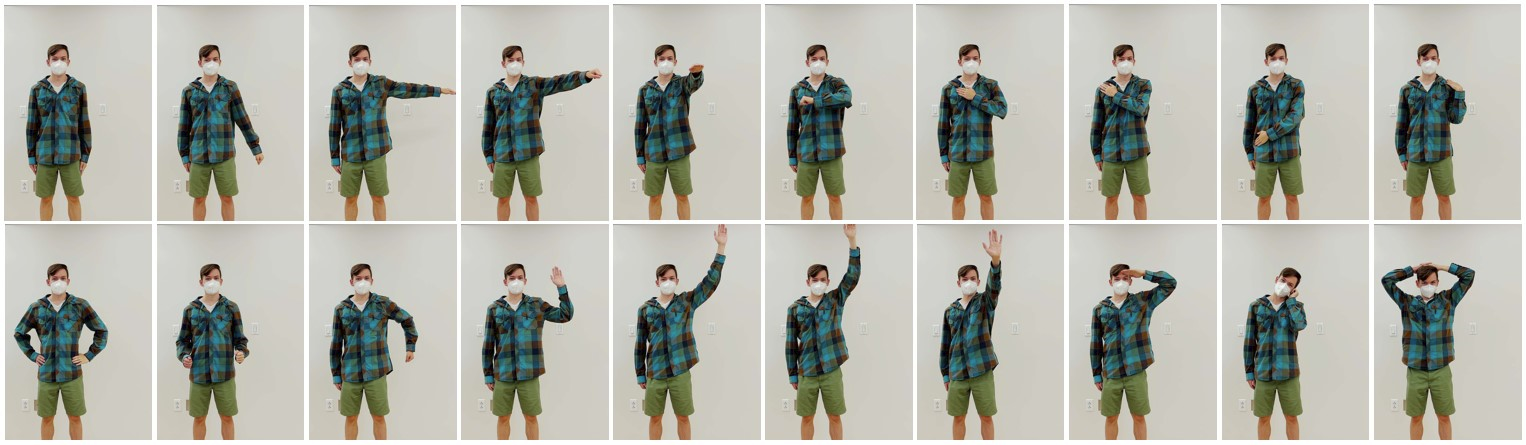}
\end{center}
  \caption{The 20 terminal arm poses utilized in our user study. Note that participants were asked to transition between these poses and all intermediate data was captured for evaluation (i.e., not just the terminal poses). }
\label{fig:arm_poses}
\Description[Photos of the terminal arm poses used in our user study.]{A series of 20 images showing the terminal arm poses in our user study. The poses range from simple gestures with the arm to the side to more dynamic gestures involving reaching above the head or crossing the arm over the body.}
\end{figure*}

\subsection{Model Training}
In summary, SmartPoser consists of two machine learning elements -- UWB correction and then pose estimation. We train the full system end-to-end using the Adam optimizer, with a learning rate of \hl{$3\mathrm{e}{-4}$}. The UWB corrector is trained to regress to the Kinect-derived distances (Section \ref{sec:uwb_acc}), scaled by user arm span, and using mean squared error loss. The pose estimator network is trained to regress to arm joint positions (shoulder, elbow, and wrist) using a mean-per-joint-position-error (MPJPE) loss. Kinect-derived ground truth arm joint positions are preprocessed to have a \hl{shoulder} frame of reference (i.e., ignoring world body rotation).  

The total loss of our system is the sum of the UWB corrector loss plus the pose estimator loss. The system is implemented using PyTorch and the PyTorch Lightning frameworks. For efficient training, we train the model to predict a full motion window of 125 samples (5 seconds) rather than just the last frame, following the method in \cite{yi_transpose_2021, huang_deep_2018}. \hl{In other words, during training we use non-overlapping windows, however for testing we overlap windows by all but one new frame.}

\subsection{Model Inference and SmartPoser Output}
To facilitate development and debugging, UWB and IMU data is streamed to a laptop (M1 MacBook Air 2021). On this machine, our trained model takes 8.1~ms per inference (on the CPU, using pytorch), meaning we output pose predictions as fast as data arrives (25 FPS). \hl{To verify SmartPoser could run entirely on-device, we also converted our trained model to CoreML with mixed precision compute and found it took only 0.99~ms per inference on an iPhone 12 Pro}. Although we envision our arm pose data most often being used internally and unseen (discussion in Section \ref{example_uses}), we did build a basic visualization tool, seen Figures~1, \ref{fig:system_diagram}, and \ref{fig:good_arm_poses}.



\section{Evaluation}\label{sec:user_study}
To evaluate the performance of SmartPoser, we conducted a user study with 10 participants (mean age 25.4; 3 identified as female, 7 as male). The study took place in a typical indoor office environment, lasted approximately 45 minutes, and compensated \$20. The study was divided into two data collection phases. The first phase focused on arm motions a user might perform in the course of their daily activities. Phase two captured less ecologically valid, but more varied and challenging gross motor arm poses (Figure \ref{fig:arm_poses}).

Before participants wore our test devices, the smartphone and smartwatch were placed side-by-side, as they might rest in a charging dock. We use this to align their independent reference frames into a common, global frame of reference. We then asked participants to wear the smartwatch on their left wrist and to place the smartphone in their left front pocket (see Figure \ref{fig:teaser}), which is a common placement \cite{ichikawa_wheres_2005,redmayne_wheres_2017} (though we note other arrangements are possible, which we discuss in Limitations). Before each study phase began, we had users perform a T-pose with their arms (palms facing down, such that the watch face is parallel to the ground) in order to capture the relative orientations of the two devices on the body (and importantly, the same study procedure used in IMUPoser \cite{mollyn_imuposer_2023}, Transpose \cite{yi_transpose_2021}, Deep Inertial Poser \cite{huang_deep_2018}, and Physical Inertial Poser \cite{yi_physical_2022}). 

\subsection{Daily Activities Procedure}
The first phase of data collection focused on arm poses that users might assume during daily routines. To capture such data, we had participants complete an "obstacle course" style study (see e.g., \cite{bao_activity_2004, intille_acquiring_2004}). We selected 20 brief mock activities, setup at stations within our study environment. These activities were divided into seven categories: \textit{Domestic} (ironing, open jar, vacuum floor, wipe table), \textit{Grooming} (comb hair, wipe forehead, wash with loofa), \textit{Food \& Beverage} (stir pot, pour teapot, drink from cup), \textit{Moving Items} (move grocery bag, lift box above head, wear bag on shoulder), \textit{Exercise} (jumping jacks, biceps curls), \textit{Home DIY} (use screwdriver, use paint roller on wall), and \textit{Miscellaneous} (check time on watch, read magazine, flick light switch). \hl{Although this set of activities is not comprehensive, we have confidence our system did not overfit. For example, the Demo Activities segment of our Video Figure shows predictions of unseen poses resulting from naturalistic activities. This is a good signal of expected performance for poses and activities not used in training and matches our anecdotal observations throughout the course of development. }

A large TV screen was positioned in the study area with a list of the activities. In addition to this visual prompt, the experimenter also spoke aloud the next activity to perform, and controlled the visual display. Although some activities were bimanual, in general we asked participants to use their left arm, as it was the one being tracked. All activities were performed one after another while data was continuously recorded, thus beyond the activities themselves, we also captured the transitions between them and miscellaneous organic gestures such as participants adjusting their clothes. Participants completed this procedure twice, taking roughly 10-15 minutes in total.

\subsection{Gross Motor Arm Pose Procedure}
Real-world activities tend to be performed in front of the body, or at least close to the torso. As such, they tend to have less variance and less extreme arm positions. Thus, as a compliment to our phase one data collection, we also captured a more gross-motor-oriented arm pose dataset. For this, we use a set of 20 terminal arm poses, seen in Figure \ref{fig:arm_poses}. These exemplify a variety of spatial endpoints, joint positions, and joint orientations. 

As with phase one, we used a TV screen to prompt participants with a random sequence of terminal poses (20 poses, 10 repeats, random order). Importantly, data was continuously captured as participants transitioned between these poses. Thus, instead of capturing 20 poses, our study captures transitions between  \textasciitilde$_{20} C _2$ pose pairs, \hl{so the effective variety and volume of motions we captured was large. In addition, the amount of time spent in the terminal poses is very low, avoiding bias.} Transitioning through the 200-pose sequence took approximately 10-15 minutes.

\subsection{Data Capture \& Ground Truth}
Throughout both phases of the study, we used an Azure Kinect \cite{tesych_about_2020} to capture ground truth data on joint positions and orientations of participants' arms. The Kinect ran continuously, and whenever we got a new frame of UWB+IMU sensor data (at \textasciitilde 25 Hz), we paired it with the most recent Kinect frame.
Across all participants, our activities dataset contained 151k frames of data (representing roughly 105 minutes of data), and our gross motor dataset contained 129K frames of data (\textasciitilde 90 minutes).

\section{Results \& Discussion}

\subsection{Train/Test Procedure}
We evaluated our model using a leave-one-participant-out cross validation scheme (i.e., ten-fold cross validation with whole participants' data as folds). This simulates "out-of-the-box" accuracy that requires only a single calibratory pose and no training data from the wearer (whereas much prior work requires per-user training data, or even per-worn-session training data). While we train on real-world activities and gross motor arm motions datasets combined, we break these results out where appropriate to give further insight into real-world accuracy. 

\subsection{Pose Tracking Accuracy}
\begin{figure}[t]
\begin{center}
 \includegraphics[width=\linewidth]{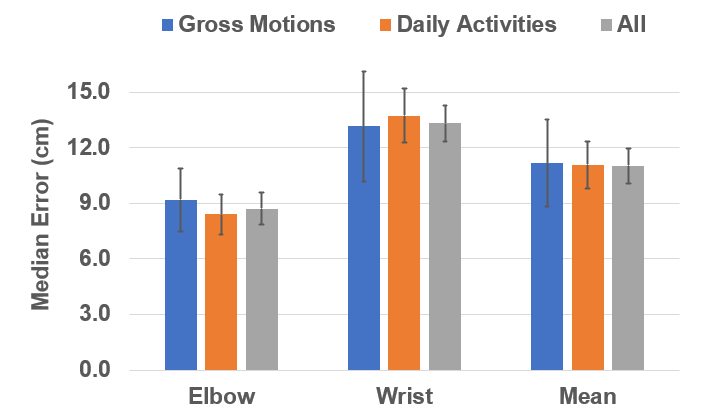}
\end{center}
  \caption{\hl{Median per-joint positional error was 11.0~cm averaged across all study participants for the elbow and wrist joints. For the gross motions phase, median per-joint positional error was 11.2~cm, 9.2~cm for just the elbow, and 13.2~cm for the wrist. For the real-world activities phase, median per-joint position error was 11.1~cm, 8.4~cm for just the elbow, and 13.7 for the wrist}. Error bars are standard deviation (cm).}
\label{fig:study_results}
\Description[Study results of median per-joint positional error.]{A bar chart showing the per-joint position error averaged across all study participants for the three arm joints, alongside mean joint error. For each joint, we have broken out the error into error for gross motions, error for daily activities, and combined error. Median joint error was highest for the wrist and lowest for the shoulder. Error also tended to be slightly higher for gross motions as opposed to daily activities.}
\end{figure}

Across all participants, we found a \hl{median error (averaging elbow and wrist joints) of 11.0~cm (SD=0.9~cm)}. As we move along the kinematic chain, error unsurprisingly increases --- median error by joint: shoulder \hl{0.17~cm (SD=0.02~cm), elbow 8.7~cm (SD=0.9~cm), and wrist 13.3~cm (SD=0.98~cm)}. Looking at error broken out by gross motor arm poses vs. real-world activities, we find \hl{median error (elbow and wrist) of 11.2~cm (SD=2.3~cm) and 11.1~cm (SD=1.3~cm) respectively.} Figure \ref{fig:study_results} provides a summary of these results for \hl{median error. Other results metrics for the combined data set of gross motions and real-world activities can be found in Table} \ref{fig:metric_table}, and the calculated CDFs in Figure \ref{fig:cdf_plot}. \hl{Our results already used a leave-one-user-out cross validation and had a low STD of 0.9 cm across users. However, to further verify we were not overfitting, we also computed metrics testing on the training set specifically and observed a median wrist error of 11.9 cm (vs. 13.3 cm from cross validation).}

\begin{table}[t]
\begin{center}
 \includegraphics[width=\linewidth]{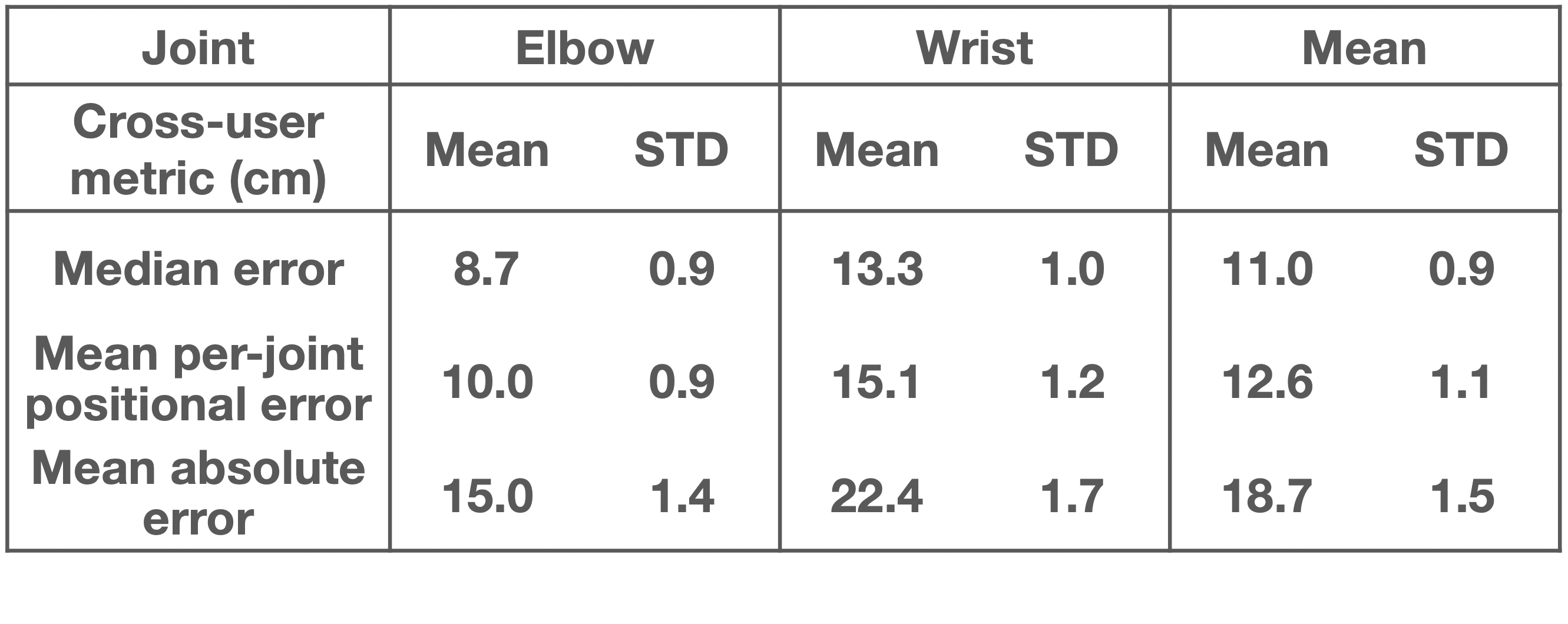}
\end{center}
  \caption{\hl{Calculated metrics for all arm joints for the combined data set of gross motions and daily activities. All metrics are reported in cm.}}
\label{fig:metric_table}
\Description[Table of results metrics]{Different metrics are shown for median error, mean per-joint positional error, and mean absolute error. Results are calculated for the combined data set and reported with both the mean and standard deviation across the 10 participants from the user study.}
\end{table}

\begin{figure}[t]
\begin{center}
 \includegraphics[width=\linewidth]{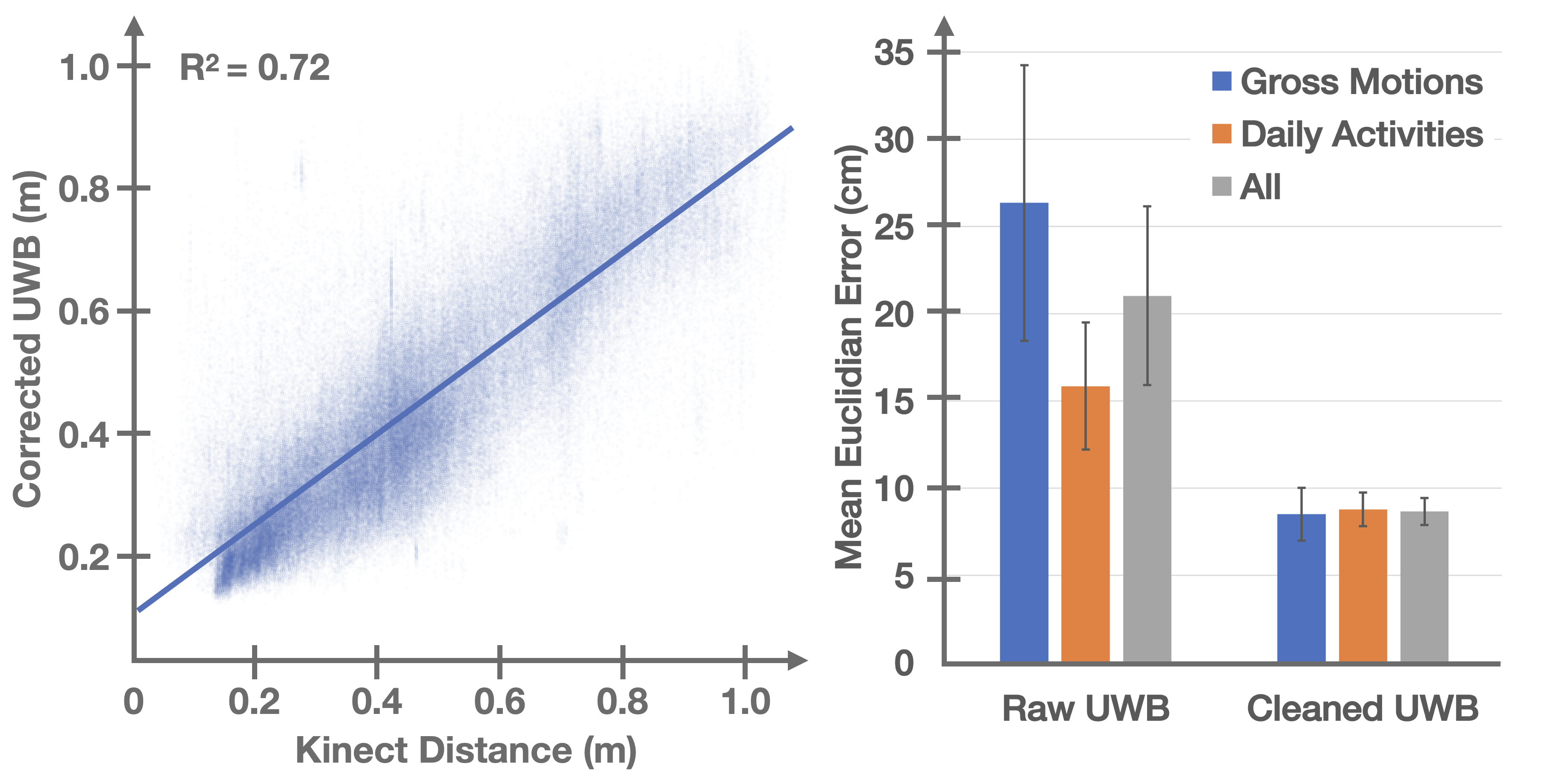}
\end{center}
  \caption{Left: Plot of corrected UWB distance vs. Kinect-derived distance (280k datapoints). Right: UWB Correction RNN performance across study datasets. Mean Euclidean error reduces from \hl{21.0}~cm to \hl{8.6}~cm on average. Error bars are standard deviation. }
\label{fig:uwb_error}
\Description[Results of the positive effect of using cleaned UWB data.]{Two plots. On the left is a line chart showing the relationship between ground truth Kinect distance and the corrected UWB distance. The R^2 correlation between the two was 0.76. On the right is a bar chart of the mean Euclidean error of distance for each motion type (gross motions, daily activities, and all) shown for using UWB data and then for cleaned UWB data. When using the cleaned UWB data, mean error reduces from 19.93 cm down to 8.54 cm on average.}
\end{figure}

\subsection{Spatial Error}
As discussed earlier, UWB ranging is sensitive to occlusion (in our case, body occlusion). Our UWB Correction RNN (Section \ref{sec:uwb_acc}) compensates for some of this error, but not entirely. Figure \ref{fig:uwb_error} compares mean euclidean error of the raw UWB distances (\hl{21.0}~cm) vs. our corrected UWB values (\hl{8.6~cm}), broken out by study dataset. We also generated a scatter plot (Figure \ref{fig:uwb_error}) plotting our corrected UWB distance vs. the Kinect-derived ground truth distance for all 131k frames of study data from our gross arm pose procedure (representing \textasciitilde 1.5 hours of data). We include a line of best fit, which has an ${R^2}=0.72$. 

Although the correlation is strong, there are other error modes and behaviors visible. To further investigate this, we created a series of heatmaps visualizing the maximum UWB error in space around the body using our study data. We look at arm pose data from all participants combined, and then only the wrist joint, which had the largest addressable volume and largest \hl{median error}. 

Figure \ref{fig:heatmaps} provides three heatmaps -- a frontal, side, and top-down view of a wearer. \hl{Heatmaps are shown for the error of raw UWB distance, corrected UWB distance, and joint predictions versus Kinect-derived ground truth. All the heatmaps have some common error hot spots (red-orange regions in Figure} \ref{fig:heatmaps}), \hl{while the volume in the front and to the left side of the body is reasonably accurate. Most of the regions show up in blue, corresponding to an error of between 0-15 cm. Wrist locations that could lead to potential self-occlusion, such as the right thigh and the space above the left shoulder, intuitively show up as highly erroneous regions in the raw UWB distance heatmaps. Note how our UWB correction module is able to account for most of these errors. The error-prone areas for the wrist joint error map well with the error-prone regions for corrected UWB distance and are altogether much much lower than the original errors for raw UWB distance. }


\begin{figure}[t]
    \centering
    \includegraphics[width=\linewidth]{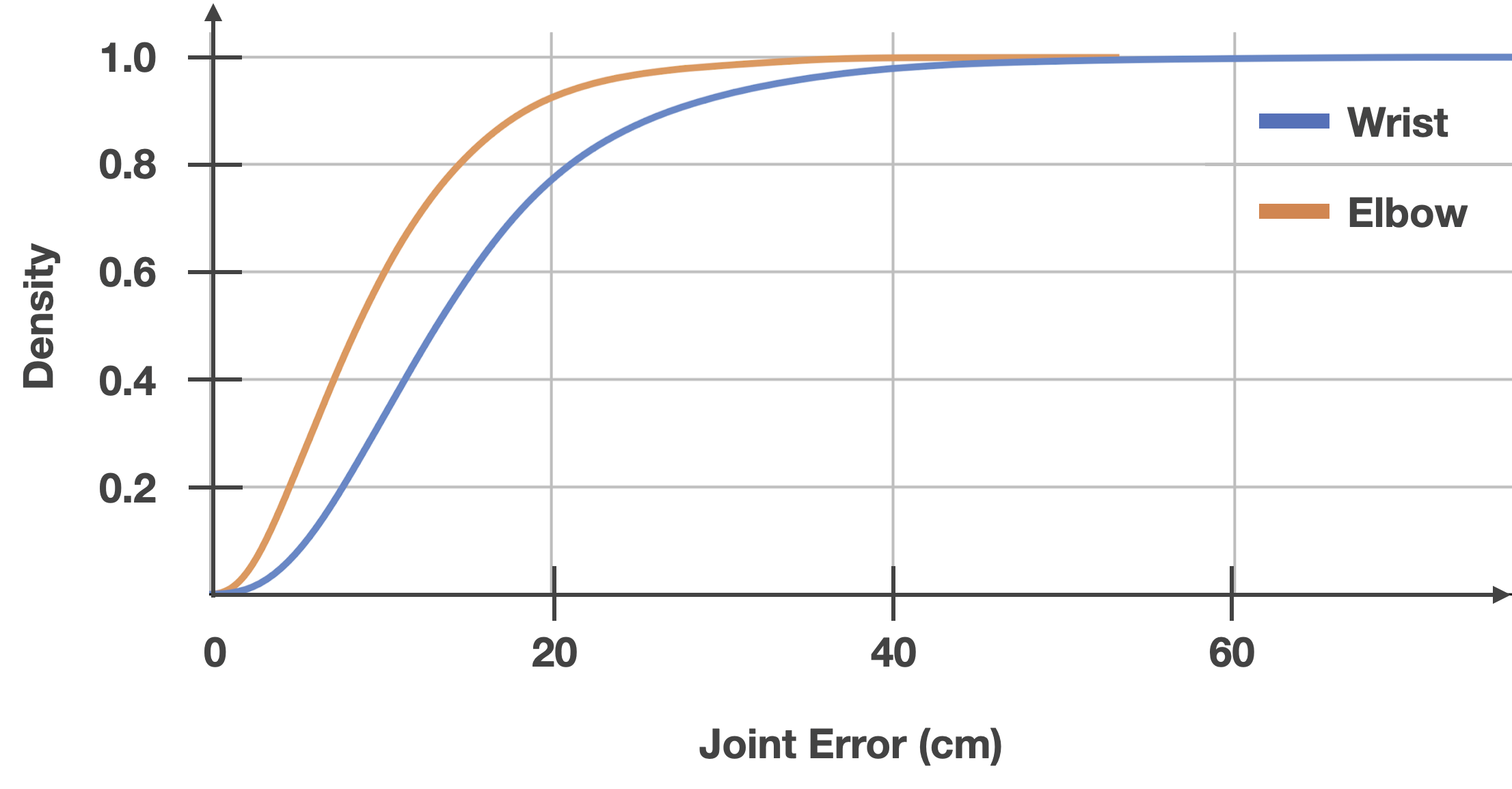}
    \caption{\hl{Cumulative distribution functions (CDF) of the elbow and wrist joint errors.}}
    \Description[Cumulative distribution functions (CDF) plots]{Cumulative distribution functions (CDF) of the elbow and wrist joint errors. Slope is higher for the elbow joint plot than the wrist joint plot.}
    \label{fig:cdf_plot}
\end{figure}

\begin{figure}[t]
\begin{center}
 \includegraphics[width=\linewidth]{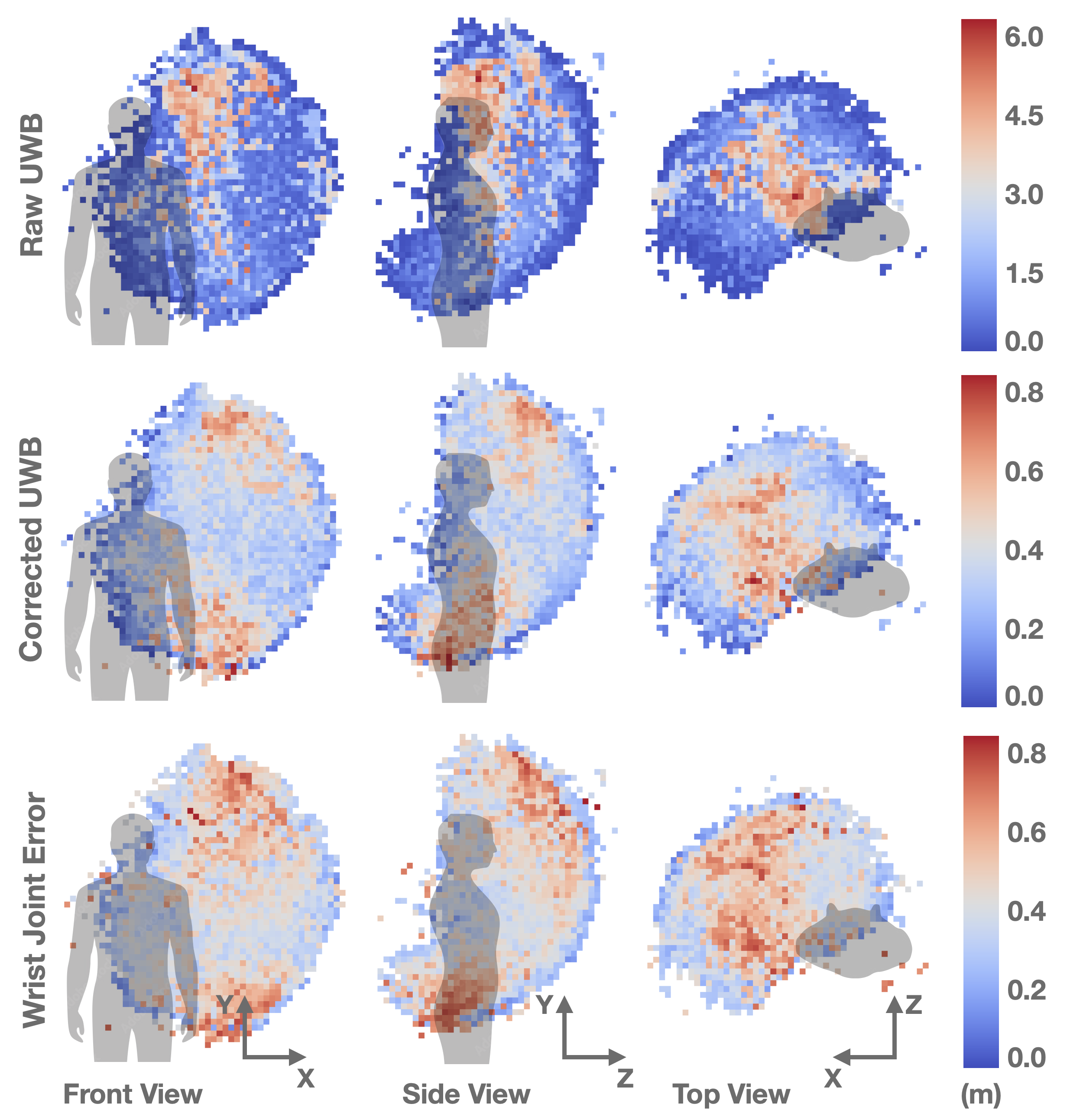}
\end{center}
  \caption{\hl{Three rows of heatmaps visualizing maximum absolute Euclidean distance error for raw UWB distances (i.e., not corrected), corrected UWB distances, and predicted wrist joint position. For all measurements, error (in meters) is calculated with respect to Kinect-derived ground truth. Note the scale for the raw UWB distances (top row) is different than the bottom two rows. }}
  \Description[Heatmaps visualizing error from the system.]{Three rows of heatmaps visualizing maximum absolute Euclidean distance error for raw UWB distances (i.e., not corrected), corrected UWB distances, and predicted wrist joint position. For all measurements, error (in meters) is calculated with respect to Kinect-derived ground truth.}
\label{fig:heatmaps}
\end{figure}

\subsection{UWB Ablation Study}

\begin{figure}[t]
    \centering
    \includegraphics[width=\linewidth]{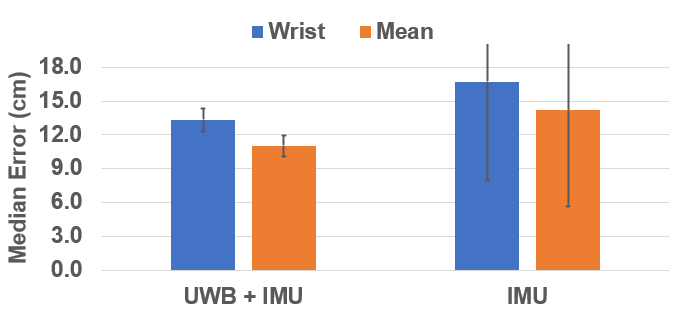}
    \caption{\hl{Ablation results showing the effectiveness using both UWB and IMU data for pose estimation. Results are calculated on data from both phases of the user study. When adding UWB data to the model for IMU alone, median error for both joints decreases from 14.2~cm to 11.0~cm.}}
    \Description[UWB ablation study results.]{Ablation results showing the effectiveness using both UWB and IMU data for pose estimation. When adding UWB data to the model for IMU alone, median error for both joints decreases from 14.2~cm to 11.0~cm.}
    \label{fig:uwb_ablation_study}
\end{figure}

\hl{The primary contribution of SmartPoser is the introduction of UWB distance data to improve arm tracking results. To quantify how much the improvement was, we ran an ablation study of our complete UWB+IMU system versus one that used only IMU data. The results of this study can be seen in Figure }\ref{fig:uwb_ablation_study}. 

\hl{Averaging the results from both the gross motions and daily activities phases, we observed a median error of 11.7~cm for the elbow and 16.7~cm for the wrist for the model with only IMU data. Compared to our standard results from the model with UWB+IMU, this meant error was 3.0~cm (25.6\%) greater for the elbow and 3.4~cm (20.4\%) greater for the wrist when using only IMU. Additionally, when looking at pose outputs, we find UWB+IMU is much more stable. For example, when holding static poses with IMUs alone they tend to "droop" slowly to a mean pose. These results show that the contribution of UWB distance measurements is significant to SmartPoser and a key to helping enable higher-fidelity pose tracking.}

\subsection{Comparison to Prior Work}
\label{comp_prior}
To contextualize our results, we compare SmartPoser accuracy to other arm tracking systems that also do not require special instrumentation  (i.e., no new or exotic sensors worn on the body) or external infrastructure (i.e., entirely self-reliant and work on-the-go). While these systems were evaluated on different datasets, we believe it is applicable to compare them to SmartPoser. 

\hl{ArmTrak} \cite{shen_i_2016} achieved a real-time median error of 12.0 cm for elbow and 13.3 cm for wrist while performing free motions with the arm (accuracy was higher when using an offline model). Of note, ArmTrak required participants to stand still (i.e., feet planted without the ability to change body orientation), unlike SmartPoser. IMUPoser \cite{mollyn_imuposer_2023} \hl{did not report median error, but achieved a combined real-time mean error of 21.6 cm for all arm joints and 26.9 cm for just the wrist. MUSE} \cite{shen_muse} \hl{only estimated position at the wrist for 5 participants and achieved a real-time (10 Hz on a desktop computer) median error of 13.0 cm for motions when the body moves around. Compared to these works, our system achieved a real-time (25 Hz) median error of 11.0~cm for the shoulder and wrist, and 13.3~cm for just the wrist.}



\section{Example Uses} \label{example_uses}

Arm pose tracking (and body pose more broadly) has been well-motivated in prior work. Our system does not enable any new applications per se, but rather makes them more practical --- able to run on off-the-shelf consumer hardware many people already own. For this reason, we do not re-implement any application demos, and instead point to the literature for motivating use cases. Please see our Video Figure for real-time demos.

\textbf{\textit{Fitness}} - Arms are involved in many exercises, and thus tracking the arms has obvious and immediate utility for fitness-oriented applications. The most straightforward are applications that automatically track rep counts for lifting free weights, jumping jacks, jump rope, and similar activities. One could also imagine a smartwatch assistant verbally guiding a user through a tai chi routine, automatically advancing steps. It may even be possible to track exercise form for personalized feedback; for example, proper extension of the arm when hitting a punching bag or form when shooting a basketball. 

\textbf{\textit{Rehabilitation}} - Tracking of the arms could also prove useful in rehabilitation (e.g., rotor cuff injuries). For example, continuous background tracking could supplement periodic sessions with physical and occupational therapists by monitoring arm metrics such as total motion, duration lifted, maximum extension, and maximum height reached. It could even be that appointments are automatically scheduled if there is limited progress or performance loss in a given metric. Alternatively, instead of passive, background monitoring, arm tracking could be employed in interactive rehabilitation applications. For example, an app could guide users through a series of balance exercises or reaching tasks. Such apps could not only automatically advance steps, but provide instructive feedback and encouragement. 

\textbf{\textit{Life Logging}} - Arm tracking could allow for new types of activities to be automatically tracked for life logging purposes. For example, it might be possible to track cigarette consumption (repeatedly raising one's hands to the mouth with a particular cadence) as part of a smoking cessation effort. Likewise, detecting periods of food consumption may also be possible for dieting apps. It may also be possible to log domestic activities, such as cooking, cleaning, vacuuming, folding clothes, and putting away dishes.

\textbf{\textit{Occupational Safety \& Training}} - Many forms of manual labor employ the hands, such as carrying boxes, stacking groceries, picking vegetables, driving a truck, laying bricks, mowing lawns, painting a house, cleaning windows, etc., which could be tracked to ensure workers are not fatigued, exploited or injured, e.g., Repetitive Strain Injury (RSI) \cite{helliwell_repetitive_2004}, Hand-Arm Vibration Syndrome (HAVS) \cite{shen_hand-arm_2017}. It may even be possible to detect improper handling of power tools or heavy equipment to improve site safety. Training on how to properly handle a tool could also happen via a companion smartwatch app. 

\textbf{\textit{Gaming}} - The commercial success of titles like Nintendo Wii Sports (the fourth best-selling video game of all time \cite{noauthor_list_2022-1}) and XBox's Dance Central franchise ushered in a new era of more physically-oriented games. While some mobile games have taken advantage of inertial or spatial data, none have leveraged arm pose to date. In particular, this could be an interesting opportunity to make smartwatch games larger and more interesting than their diminutive screens might otherwise allow. For instance, users could practice their golf swings or frisbee throws, with the watch reporting on achieved range and form quality. While we only considered arm pose tracking with the phone stored in the pocket, our technique should also extend to a phone held in the hand (with a different machine learning model). This would allow for gaming experiences with coordinated graphics. More specifically, the user would look at a game running on the phone held in one hand, with the other hand free to perform game actions, such as throwing a ball or fighting with a sword. 

\textbf{\textit{Context-Aware Interfaces \& Assistants}} - Devices that are context-aware can automatically summon relevant interfaces and information to augment a user's activity. Arm pose is a useful piece of information that could improve the accuracy and breadth of prior context sensing work (using e.g., environmental sounds \cite{laput_ubicoustics_2018}, bioacoustics \cite{laput_viband_2016}, movement data \cite{laput_sensing_2019}). For instance, arm pose tracking could help disambiguate if a user is driving an automobile vs. being a passenger. As discussed in the above subsections on Life Logging and Occupational Safety and Training, there is a long tail of arm-centric activities that could be automatically detected, which could launch context-aware functionality.

\section{Open Source}

To enable others to build upon our system, we have made our dataset, architecture, and trained model freely available at\break \url{https://github.com/FIGLAB/SmartPoser}, with the gracious permission of our participants.

\section{Limitations \& Future Work} \label{limitations}

We believe our proof-of-concept implementation demonstrates the feasibility of our method, but there are nevertheless limitations of note. First among these limitations is that our current work only considers the case when the phone is stored in a front pocket, a common but not ubiquitous placement \cite{ichikawa_wheres_2005, redmayne_wheres_2017}. In general, our pipeline should scale to other pocket locations, including rear pants pockets, and potentially even cycling jersey pockets and fanny packs. However, when the phone is taken out and held in the hand, and can exist in many possible positions with respect to the body, a third spatial reference would be required (potentially head position captured via the phone's user-facing camera or earbuds with UWB). There already exists work that can automatically detect pocket vs. hand phone placement \cite{wiese_phoneprioception_2013, harrison_lightweight_2008}, which could toggle between different models. 

Another piece of information we need is which pocket a user typically stores their phone and which wrist the user wears a watch, which could be captured during a one-time setup wizard. Even with all this data, the user still needs to perform a calibration procedure, which is a limitation of our current implementation. This process is described in the Evaluation section, but briefly here, participants in our study had to hold the devices near each other, \hl{then assume a T-pose with their arm to orient the IMU reference frames} (e.g. to offset how the phone is placed rotationally in the pocket). While inconvenient to perform once, we found the system was able to maintain its calibration for an extended period of time. \hl{We note the T-pose used is not special and could be replaced by a pose users more naturally perform, such as arm by side (potentially auto-detected using UWB distance). If the calibration process could be condensed to a single gesture, such as holding the arms by the side of the body}, we imagine there are many contexts in which it would be acceptable to do this once at the beginning of an activity, such as starting a workout or a rehab session. This is one aspect where the enhanced tracking mode for the iOS UWB API mentioned in the Implementation section could be particularly useful. This mode provides both distance and 3D angles between the phone and watch. Thus, if the watch was held in the phone's FoV cone, the phone would know its absolute position relative to the watch and could use this to use this data to calibrate the IMU reference frames automatically. 

One additional issue is that many people do not carry their phone in their pocket at all \cite{ichikawa_wheres_2005, redmayne_wheres_2017}. Feminine clothing lacking functional pockets \cite{spencer_pocket_2020} is a prominent example. People may opt to carry their phone in a bag or purse. Tracking watch position with a phone in a bag is a more challenging problem than a phone in a pocket because, while there is variation in pocket design, the phone is located in the same area on the body in the majority of cases. This is not true of bags where the space of possible phone locations is much larger. \hl{Since the phone could move inside bags, we do not expect SmartPoser to work well in such situations. However, as a solution, instead of using a phone, our system could generalize to other devices that are worn in static locations and may become UWB-enabled in the future, such as smart glasses or earbuds.}

\section{Conclusion}
We have presented SmartPoser, a system that enables 3D arm pose tracking using an off-the-shelf smartphone and smartwatch in their normally-worn positions. In our approach, we build upon a state-of-the-art bidirectional LSTM model for arm pose prediction and fused more traditional inertial data with new UWB time-of-flight range measurements between the two devices. Independent of using commodity devices, this technique is novel as UWB ranging has never been used to augment IMU data in a wearable pose-tracking system before. Through a user study involving real-world activities and gross arm motions, our approach yielded a \hl{median positional error of 11.0~cm for the wrist and elbow joints, with no user data in the training set}. This is a level of tracking fidelity that has only been achieved in the past through full-body instrumentation with IMUs or UWB-instrumented environments. Overall, by achieving this result in this consumer form factor, we believe SmartPoser provides the most realistic pathway to date for arm-based tracking and interactions to be adopted for widespread use.

\begin{acks}
We would like to thank Riku Arakawa for his early help with application development. We are also grateful to Dr. David Lindlbauer and the Augmented Perception Lab at Carnegie Mellon University for sharing technical equipment and support.
\end{acks}

\bibliographystyle{ACM-Reference-Format}
\bibliography{references}


\begin{thebibliography}{63}


\ifx \showCODEN    \undefined \def \showCODEN     #1{\unskip}     \fi
\ifx \showDOI      \undefined \def \showDOI       #1{#1}\fi
\ifx \showISBNx    \undefined \def \showISBNx     #1{\unskip}     \fi
\ifx \showISBNxiii \undefined \def \showISBNxiii  #1{\unskip}     \fi
\ifx \showISSN     \undefined \def \showISSN      #1{\unskip}     \fi
\ifx \showLCCN     \undefined \def \showLCCN      #1{\unskip}     \fi
\ifx \shownote     \undefined \def \shownote      #1{#1}          \fi
\ifx \showarticletitle \undefined \def \showarticletitle #1{#1}   \fi
\ifx \showURL      \undefined \def \showURL       {\relax}        \fi
\providecommand\bibfield[2]{#2}
\providecommand\bibinfo[2]{#2}
\providecommand\natexlab[1]{#1}
\providecommand\showeprint[2][]{arXiv:#2}

\bibitem[Apple(2022a)]%
        {noauthor_core_2022}
\bibfield{author}{\bibinfo{person}{Apple}.} \bibinfo{year}{2022}\natexlab{a}.
\newblock \bibinfo{title}{Core {Motion} {\textbar} {Apple} {Developer}
  {Documentation}}.
\newblock
\newblock
\urldef\tempurl%
\url{https://developer.apple.com/documentation/coremotion/}
\showURL{%
\tempurl}


\bibitem[Apple(2022b)]%
        {noauthor_nearby_2022}
\bibfield{author}{\bibinfo{person}{Apple}.} \bibinfo{year}{2022}\natexlab{b}.
\newblock \bibinfo{title}{Nearby {Interaction} {\textbar} {Apple} {Developer}
  {Documentation}}.
\newblock
\newblock
\urldef\tempurl%
\url{https://developer.apple.com/documentation/nearbyinteraction}
\showURL{%
\tempurl}


\bibitem[Apple(2023)]%
        {handwashingApple}
\bibfield{author}{\bibinfo{person}{Apple}.} \bibinfo{year}{2023}\natexlab{}.
\newblock \bibinfo{title}{Set up Handwashing on Apple Watch}.
\newblock
\newblock
\urldef\tempurl%
\url{https://support.apple.com/guide/watch/set-up-handwashing-apdc9b9f04a8/watchos}
\showURL{%
\tempurl}


\bibitem[Bachmann et~al\mbox{.}(2001)]%
        {bachmann_inertial_2001}
\bibfield{author}{\bibinfo{person}{Eric~R. Bachmann},
  \bibinfo{person}{Robert~B. McGhee}, \bibinfo{person}{Xiaoping Yun}, {and}
  \bibinfo{person}{Michael~J. Zyda}.} \bibinfo{year}{2001}\natexlab{}.
\newblock \showarticletitle{Inertial and magnetic posture tracking for
  inserting humans into networked virtual environments}. In
  \bibinfo{booktitle}{\emph{Proceedings of the {ACM} symposium on {Virtual}
  reality software and technology}} \emph{(\bibinfo{series}{{VRST} '01})}.
  \bibinfo{publisher}{Association for Computing Machinery},
  \bibinfo{address}{New York, NY, USA}, \bibinfo{pages}{9--16}.
\newblock
\showISBNx{978-1-58113-427-8}
\urldef\tempurl%
\url{https://doi.org/10.1145/505008.505011}
\showDOI{\tempurl}


\bibitem[Baldominos et~al\mbox{.}(2015)]%
        {baldominos_approach_2015}
\bibfield{author}{\bibinfo{person}{Alejandro Baldominos}, \bibinfo{person}{Yago
  Saez}, {and} \bibinfo{person}{Cristina García~del Pozo}.}
  \bibinfo{year}{2015}\natexlab{}.
\newblock \showarticletitle{An {Approach} to {Physical} {Rehabilitation}
  {Using} {State}-of-the-art {Virtual} {Reality} and {Motion} {Tracking}
  {Technologies}}.
\newblock \bibinfo{journal}{\emph{Procedia Computer Science}}
  \bibinfo{volume}{64} (\bibinfo{year}{2015}), \bibinfo{pages}{10--16}.
\newblock
\showISSN{18770509}
\urldef\tempurl%
\url{https://doi.org/10.1016/j.procs.2015.08.457}
\showDOI{\tempurl}


\bibitem[Bao and Intille(2004)]%
        {bao_activity_2004}
\bibfield{author}{\bibinfo{person}{Ling Bao} {and} \bibinfo{person}{Stephen~S.
  Intille}.} \bibinfo{year}{2004}\natexlab{}.
\newblock \showarticletitle{Activity {Recognition} from {User}-{Annotated}
  {Acceleration} {Data}}. In \bibinfo{booktitle}{\emph{Pervasive {Computing}}}
  \emph{(\bibinfo{series}{Lecture {Notes} in {Computer} {Science}})},
  \bibfield{editor}{\bibinfo{person}{Alois Ferscha} {and}
  \bibinfo{person}{Friedemann Mattern}} (Eds.). \bibinfo{publisher}{Springer},
  \bibinfo{address}{Berlin, Heidelberg}, \bibinfo{pages}{1--17}.
\newblock
\showISBNx{978-3-540-24646-6}
\urldef\tempurl%
\url{https://doi.org/10.1007/978-3-540-24646-6_1}
\showDOI{\tempurl}


\bibitem[Bosché et~al\mbox{.}(2016)]%
        {bosche_towards_2016}
\bibfield{author}{\bibinfo{person}{Frédéric Bosché},
  \bibinfo{person}{Mohamed Abdel-Wahab}, {and} \bibinfo{person}{Ludovico
  Carozza}.} \bibinfo{year}{2016}\natexlab{}.
\newblock \showarticletitle{Towards a {Mixed} {Reality} {System} for
  {Construction} {Trade} {Training}}.
\newblock \bibinfo{journal}{\emph{Journal of Computing in Civil Engineering}}
  \bibinfo{volume}{30}, \bibinfo{number}{2} (\bibinfo{date}{March}
  \bibinfo{year}{2016}), \bibinfo{pages}{04015016}.
\newblock
\showISSN{1943-5487}
\urldef\tempurl%
\url{https://doi.org/10.1061/(ASCE)CP.1943-5487.0000479}
\showDOI{\tempurl}
\newblock
\shownote{Publisher: American Society of Civil Engineers}.


\bibitem[Butt et~al\mbox{.}(2021)]%
        {butt_magnetometer_2021}
\bibfield{author}{\bibinfo{person}{Hammad~Tanveer Butt},
  \bibinfo{person}{Bertram Taetz}, \bibinfo{person}{Mathias Musahl},
  \bibinfo{person}{Maria~A. Sanchez}, \bibinfo{person}{Pramod Murthy}, {and}
  \bibinfo{person}{Didier Stricker}.} \bibinfo{year}{2021}\natexlab{}.
\newblock \showarticletitle{Magnetometer {Robust} {Deep} {Human} {Pose}
  {Regression} {With} {Uncertainty} {Prediction} {Using} {Sparse} {Body} {Worn}
  {Magnetic} {Inertial} {Measurement} {Units}}.
\newblock \bibinfo{journal}{\emph{IEEE Access}}  \bibinfo{volume}{9}
  (\bibinfo{year}{2021}), \bibinfo{pages}{36657--36673}.
\newblock
\showISSN{2169-3536}
\urldef\tempurl%
\url{https://doi.org/10.1109/ACCESS.2021.3062545}
\showDOI{\tempurl}
\newblock
\shownote{Conference Name: IEEE Access}.


\bibitem[Corrales et~al\mbox{.}(2008)]%
        {corrales_hybrid_2008}
\bibfield{author}{\bibinfo{person}{J.~A. Corrales}, \bibinfo{person}{F.~A.
  Candelas}, {and} \bibinfo{person}{F. Torres}.}
  \bibinfo{year}{2008}\natexlab{}.
\newblock \showarticletitle{Hybrid tracking of human operators using
  {IMU}/{UWB} data fusion by a {Kalman} filter}. In
  \bibinfo{booktitle}{\emph{2008 3rd {ACM}/{IEEE} {International} {Conference}
  on {Human}-{Robot} {Interaction} ({HRI})}}. \bibinfo{pages}{193--200}.
\newblock
\urldef\tempurl%
\url{https://doi.org/10.1145/1349822.1349848}
\showDOI{\tempurl}
\newblock
\shownote{ISSN: 2167-2148}.


\bibitem[Ding et~al\mbox{.}(2017)]%
        {ding_platform_2017}
\bibfield{author}{\bibinfo{person}{Han Ding}, \bibinfo{person}{Jinsong Han},
  \bibinfo{person}{Longfei Shangguan}, \bibinfo{person}{Wei Xi},
  \bibinfo{person}{Zhiping Jiang}, \bibinfo{person}{Zheng Yang},
  \bibinfo{person}{Zimu Zhou}, \bibinfo{person}{Panlong Yang}, {and}
  \bibinfo{person}{Jizhong Zhao}.} \bibinfo{year}{2017}\natexlab{}.
\newblock \showarticletitle{A {Platform} for {Free}-{Weight} {Exercise}
  {Monitoring} with {Passive} {Tags}}.
\newblock \bibinfo{journal}{\emph{IEEE Transactions on Mobile Computing}}
  \bibinfo{volume}{16}, \bibinfo{number}{12} (\bibinfo{date}{Dec.}
  \bibinfo{year}{2017}), \bibinfo{pages}{3279--3293}.
\newblock
\showISSN{1558-0660}
\urldef\tempurl%
\url{https://doi.org/10.1109/TMC.2017.2691705}
\showDOI{\tempurl}
\newblock
\shownote{Conference Name: IEEE Transactions on Mobile Computing}.


\bibitem[Google(2022)]%
        {google_uwb}
\bibfield{author}{\bibinfo{person}{Google}.} \bibinfo{year}{2022}\natexlab{}.
\newblock \bibinfo{title}{Android Developers, Ultra-wideband (UWB)
  communication}.
\newblock
\newblock
\urldef\tempurl%
\url{https://developer.android.com/guide/topics/connectivity/uwb}
\showURL{%
\tempurl}


\bibitem[Großwindhager et~al\mbox{.}(2018)]%
        {grobwindhager_uwb}
\bibfield{author}{\bibinfo{person}{Bernhard Großwindhager},
  \bibinfo{person}{Carlo Alberto~Boano}, \bibinfo{person}{Michael Rath}, {and}
  \bibinfo{person}{Kay Römer}.} \bibinfo{year}{2018}\natexlab{}.
\newblock \showarticletitle{Enabling Runtime Adaptation of Physical Layer
  Settings for Dependable UWB Communications}. In
  \bibinfo{booktitle}{\emph{2018 IEEE 19th International Symposium on "A World
  of Wireless, Mobile and Multimedia Networks" (WoWMoM)}}.
  \bibinfo{pages}{01--11}.
\newblock
\urldef\tempurl%
\url{https://doi.org/10.1109/WoWMoM.2018.8449776}
\showDOI{\tempurl}


\bibitem[Harrison and Hudson(2008)]%
        {harrison_lightweight_2008}
\bibfield{author}{\bibinfo{person}{Chris Harrison} {and}
  \bibinfo{person}{Scott~E. Hudson}.} \bibinfo{year}{2008}\natexlab{}.
\newblock \showarticletitle{Lightweight material detection for placement-aware
  mobile computing}. In \bibinfo{booktitle}{\emph{Proceedings of the 21st
  annual {ACM} symposium on {User} interface software and technology}}
  \emph{(\bibinfo{series}{{UIST} '08})}. \bibinfo{publisher}{Association for
  Computing Machinery}, \bibinfo{address}{New York, NY, USA},
  \bibinfo{pages}{279--282}.
\newblock
\showISBNx{978-1-59593-975-3}
\urldef\tempurl%
\url{https://doi.org/10.1145/1449715.1449761}
\showDOI{\tempurl}


\bibitem[Helliwell and Taylor(2004)]%
        {helliwell_repetitive_2004}
\bibfield{author}{\bibinfo{person}{P Helliwell} {and} \bibinfo{person}{W
  Taylor}.} \bibinfo{year}{2004}\natexlab{}.
\newblock \showarticletitle{Repetitive strain injury}.
\newblock \bibinfo{journal}{\emph{Postgraduate Medical Journal}}
  \bibinfo{volume}{80}, \bibinfo{number}{946} (\bibinfo{date}{Aug.}
  \bibinfo{year}{2004}), \bibinfo{pages}{438--443}.
\newblock
\showISSN{0032-5473}
\urldef\tempurl%
\url{https://doi.org/10.1136/pgmj.2003.012591}
\showDOI{\tempurl}


\bibitem[Hol et~al\mbox{.}(2009)]%
        {hol_tightly_2009}
\bibfield{author}{\bibinfo{person}{Jeroen~D. Hol}, \bibinfo{person}{Fred
  Dijkstra}, \bibinfo{person}{Henk Luinge}, {and} \bibinfo{person}{Thomas~B.
  Schon}.} \bibinfo{year}{2009}\natexlab{}.
\newblock \showarticletitle{Tightly coupled {UWB}/{IMU} pose estimation}. In
  \bibinfo{booktitle}{\emph{2009 {IEEE} {International} {Conference} on
  {Ultra}-{Wideband}}}. \bibinfo{pages}{688--692}.
\newblock
\urldef\tempurl%
\url{https://doi.org/10.1109/ICUWB.2009.5288724}
\showDOI{\tempurl}


\bibitem[Huang et~al\mbox{.}(2018)]%
        {huang_deep_2018}
\bibfield{author}{\bibinfo{person}{Yinghao Huang}, \bibinfo{person}{Manuel
  Kaufmann}, \bibinfo{person}{Emre Aksan}, \bibinfo{person}{Michael~J. Black},
  \bibinfo{person}{Otmar Hilliges}, {and} \bibinfo{person}{Gerard Pons-Moll}.}
  \bibinfo{year}{2018}\natexlab{}.
\newblock \showarticletitle{Deep inertial poser: learning to reconstruct human
  pose from sparse inertial measurements in real time}.
\newblock \bibinfo{journal}{\emph{ACM Transactions on Graphics}}
  \bibinfo{volume}{37}, \bibinfo{number}{6} (\bibinfo{date}{Dec.}
  \bibinfo{year}{2018}), \bibinfo{pages}{1--15}.
\newblock
\showISSN{0730-0301, 1557-7368}
\urldef\tempurl%
\url{https://doi.org/10.1145/3272127.3275108}
\showDOI{\tempurl}


\bibitem[Ichikawa et~al\mbox{.}(2005)]%
        {ichikawa_wheres_2005}
\bibfield{author}{\bibinfo{person}{Fumiko Ichikawa}, \bibinfo{person}{Jan
  Chipchase}, {and} \bibinfo{person}{R. Grignani}.}
  \bibinfo{year}{2005}\natexlab{}.
\newblock \showarticletitle{Where's the phone? {A} study of mobile phone
  location in public spaces}.
\newblock \bibinfo{journal}{\emph{Mobile Technology, Applications and Systems,
  2005 2nd International Conference on}} (\bibinfo{date}{Jan.}
  \bibinfo{year}{2005}), \bibinfo{pages}{1--8}.
\newblock


\bibitem[Intille et~al\mbox{.}(2004)]%
        {intille_acquiring_2004}
\bibfield{author}{\bibinfo{person}{Stephen~S. Intille}, \bibinfo{person}{Ling
  Bao}, \bibinfo{person}{Emmanuel~Munguia Tapia}, {and} \bibinfo{person}{John
  Rondoni}.} \bibinfo{year}{2004}\natexlab{}.
\newblock \showarticletitle{Acquiring in situ training data for context-aware
  ubiquitous computing applications}. In \bibinfo{booktitle}{\emph{Proceedings
  of the {SIGCHI} {Conference} on {Human} {Factors} in {Computing} {Systems}}}
  \emph{(\bibinfo{series}{{CHI} '04})}. \bibinfo{publisher}{Association for
  Computing Machinery}, \bibinfo{address}{New York, NY, USA},
  \bibinfo{pages}{1--8}.
\newblock
\showISBNx{978-1-58113-702-6}
\urldef\tempurl%
\url{https://doi.org/10.1145/985692.985693}
\showDOI{\tempurl}


\bibitem[Jahren et~al\mbox{.}(2021)]%
        {jahren_towards_2021}
\bibfield{author}{\bibinfo{person}{Silje~Ekroll Jahren}, \bibinfo{person}{Niels
  Aakvaag}, \bibinfo{person}{Frode Strisland}, \bibinfo{person}{Andreas Vogl},
  \bibinfo{person}{Alessandro Liberale}, {and} \bibinfo{person}{Anders~E.
  Liverud}.} \bibinfo{year}{2021}\natexlab{}.
\newblock \showarticletitle{Towards {Human} {Motion} {Tracking} {Enhanced} by
  {Semi}-{Continuous} {Ultrasonic} {Time}-of-{Flight} {Measurements}}.
\newblock \bibinfo{journal}{\emph{Sensors}} \bibinfo{volume}{21},
  \bibinfo{number}{7} (\bibinfo{date}{Jan.} \bibinfo{year}{2021}),
  \bibinfo{pages}{2259}.
\newblock
\showISSN{1424-8220}
\urldef\tempurl%
\url{https://doi.org/10.3390/s21072259}
\showDOI{\tempurl}


\bibitem[Jiang et~al\mbox{.}(2022)]%
        {jiang_avatarposer_2022}
\bibfield{author}{\bibinfo{person}{Jiaxi Jiang}, \bibinfo{person}{Paul Streli},
  \bibinfo{person}{Huajian Qiu}, \bibinfo{person}{Andreas Fender},
  \bibinfo{person}{Larissa Laich}, \bibinfo{person}{Patrick Snape}, {and}
  \bibinfo{person}{Christian Holz}.} \bibinfo{year}{2022}\natexlab{}.
\newblock \bibinfo{title}{{AvatarPoser}: {Articulated} {Full}-{Body} {Pose}
  {Tracking} from {Sparse} {Motion} {Sensing}}.
\newblock
\newblock
\urldef\tempurl%
\url{http://arxiv.org/abs/2207.13784}
\showURL{%
\tempurl}
\newblock
\shownote{arXiv:2207.13784 [cs]}.


\bibitem[Jin et~al\mbox{.}(2018)]%
        {jin_towards_2018}
\bibfield{author}{\bibinfo{person}{Haojian Jin}, \bibinfo{person}{Zhijian
  Yang}, \bibinfo{person}{Swarun Kumar}, {and} \bibinfo{person}{Jason~I.
  Hong}.} \bibinfo{year}{2018}\natexlab{}.
\newblock \showarticletitle{Towards {Wearable} {Everyday} {Body}-{Frame}
  {Tracking} using {Passive} {RFIDs}}.
\newblock \bibinfo{journal}{\emph{Proceedings of the ACM on Interactive,
  Mobile, Wearable and Ubiquitous Technologies}} \bibinfo{volume}{1},
  \bibinfo{number}{4} (\bibinfo{date}{Jan.} \bibinfo{year}{2018}),
  \bibinfo{pages}{145:1--145:23}.
\newblock
\urldef\tempurl%
\url{https://doi.org/10.1145/3161199}
\showDOI{\tempurl}


\bibitem[Laput et~al\mbox{.}(2018)]%
        {laput_ubicoustics_2018}
\bibfield{author}{\bibinfo{person}{Gierad Laput}, \bibinfo{person}{Karan
  Ahuja}, \bibinfo{person}{Mayank Goel}, {and} \bibinfo{person}{Chris
  Harrison}.} \bibinfo{year}{2018}\natexlab{}.
\newblock \showarticletitle{Ubicoustics: {Plug}-and-{Play} {Acoustic}
  {Activity} {Recognition}}. In \bibinfo{booktitle}{\emph{Proceedings of the
  31st {Annual} {ACM} {Symposium} on {User} {Interface} {Software} and
  {Technology}}} \emph{(\bibinfo{series}{{UIST} '18})}.
  \bibinfo{publisher}{Association for Computing Machinery},
  \bibinfo{address}{New York, NY, USA}, \bibinfo{pages}{213--224}.
\newblock
\showISBNx{978-1-4503-5948-1}
\urldef\tempurl%
\url{https://doi.org/10.1145/3242587.3242609}
\showDOI{\tempurl}


\bibitem[Laput and Harrison(2019)]%
        {laput_sensing_2019}
\bibfield{author}{\bibinfo{person}{Gierad Laput} {and} \bibinfo{person}{Chris
  Harrison}.} \bibinfo{year}{2019}\natexlab{}.
\newblock \showarticletitle{Sensing {Fine}-{Grained} {Hand} {Activity} with
  {Smartwatches}}. In \bibinfo{booktitle}{\emph{Proceedings of the 2019 {CHI}
  {Conference} on {Human} {Factors} in {Computing} {Systems}}}
  \emph{(\bibinfo{series}{{CHI} '19})}. \bibinfo{publisher}{Association for
  Computing Machinery}, \bibinfo{address}{New York, NY, USA},
  \bibinfo{pages}{1--13}.
\newblock
\showISBNx{978-1-4503-5970-2}
\urldef\tempurl%
\url{https://doi.org/10.1145/3290605.3300568}
\showDOI{\tempurl}


\bibitem[Laput et~al\mbox{.}(2016)]%
        {laput_viband_2016}
\bibfield{author}{\bibinfo{person}{Gierad Laput}, \bibinfo{person}{Robert
  Xiao}, {and} \bibinfo{person}{Chris Harrison}.}
  \bibinfo{year}{2016}\natexlab{}.
\newblock \showarticletitle{{ViBand}: {High}-{Fidelity} {Bio}-{Acoustic}
  {Sensing} {Using} {Commodity} {Smartwatch} {Accelerometers}}. In
  \bibinfo{booktitle}{\emph{Proceedings of the 29th {Annual} {Symposium} on
  {User} {Interface} {Software} and {Technology}}}
  \emph{(\bibinfo{series}{{UIST} '16})}. \bibinfo{publisher}{Association for
  Computing Machinery}, \bibinfo{address}{New York, NY, USA},
  \bibinfo{pages}{321--333}.
\newblock
\showISBNx{978-1-4503-4189-9}
\urldef\tempurl%
\url{https://doi.org/10.1145/2984511.2984582}
\showDOI{\tempurl}


\bibitem[Lisini~Baldi et~al\mbox{.}(2020)]%
        {lisini_baldi_upper_2020}
\bibfield{author}{\bibinfo{person}{Tommaso Lisini~Baldi},
  \bibinfo{person}{Francesco Farina}, \bibinfo{person}{Andrea Garulli},
  \bibinfo{person}{Antonio Giannitrapani}, {and} \bibinfo{person}{Domenico
  Prattichizzo}.} \bibinfo{year}{2020}\natexlab{}.
\newblock \showarticletitle{Upper {Body} {Pose} {Estimation} {Using} {Wearable}
  {Inertial} {Sensors} and {Multiplicative} {Kalman} {Filter}}.
\newblock \bibinfo{journal}{\emph{IEEE Sensors Journal}} \bibinfo{volume}{20},
  \bibinfo{number}{1} (\bibinfo{date}{Jan.} \bibinfo{year}{2020}),
  \bibinfo{pages}{492--500}.
\newblock
\showISSN{1558-1748}
\urldef\tempurl%
\url{https://doi.org/10.1109/JSEN.2019.2940612}
\showDOI{\tempurl}
\newblock
\shownote{Conference Name: IEEE Sensors Journal}.


\bibitem[Liu et~al\mbox{.}(2011)]%
        {liu_realtime_2011}
\bibfield{author}{\bibinfo{person}{Huajun Liu}, \bibinfo{person}{Xiaolin Wei},
  \bibinfo{person}{Jinxiang Chai}, \bibinfo{person}{Inwoo Ha}, {and}
  \bibinfo{person}{Taehyun Rhee}.} \bibinfo{year}{2011}\natexlab{}.
\newblock \showarticletitle{Realtime human motion control with a small number
  of inertial sensors}. In \bibinfo{booktitle}{\emph{Symposium on {Interactive}
  {3D} {Graphics} and {Games}}} \emph{(\bibinfo{series}{{I3D} '11})}.
  \bibinfo{publisher}{Association for Computing Machinery},
  \bibinfo{address}{New York, NY, USA}, \bibinfo{pages}{133--140}.
\newblock
\showISBNx{978-1-4503-0565-5}
\urldef\tempurl%
\url{https://doi.org/10.1145/1944745.1944768}
\showDOI{\tempurl}


\bibitem[Liu et~al\mbox{.}(2020)]%
        {liu_wearable_2020}
\bibfield{author}{\bibinfo{person}{Shi~Qiang Liu}, \bibinfo{person}{Jun~Chang
  Zhang}, {and} \bibinfo{person}{Rong Zhu}.} \bibinfo{year}{2020}\natexlab{}.
\newblock \showarticletitle{A {Wearable} {Human} {Motion} {Tracking} {Device}
  {Using} {Micro} {Flow} {Sensor} {Incorporating} a {Micro} {Accelerometer}}.
\newblock \bibinfo{journal}{\emph{IEEE Transactions on Biomedical Engineering}}
  \bibinfo{volume}{67}, \bibinfo{number}{4} (\bibinfo{date}{April}
  \bibinfo{year}{2020}), \bibinfo{pages}{940--948}.
\newblock
\showISSN{1558-2531}
\urldef\tempurl%
\url{https://doi.org/10.1109/TBME.2019.2924689}
\showDOI{\tempurl}
\newblock
\shownote{Conference Name: IEEE Transactions on Biomedical Engineering}.


\bibitem[Marins et~al\mbox{.}(2001)]%
        {marins_extended_2001}
\bibfield{author}{\bibinfo{person}{J.L. Marins}, \bibinfo{person}{Xiaoping
  Yun}, \bibinfo{person}{E.R. Bachmann}, \bibinfo{person}{R.B. McGhee}, {and}
  \bibinfo{person}{M.J. Zyda}.} \bibinfo{year}{2001}\natexlab{}.
\newblock \showarticletitle{An extended {Kalman} filter for quaternion-based
  orientation estimation using {MARG} sensors}. In
  \bibinfo{booktitle}{\emph{Proceedings 2001 {IEEE}/{RSJ} {International}
  {Conference} on {Intelligent} {Robots} and {Systems}. {Expanding} the
  {Societal} {Role} of {Robotics} in the the {Next} {Millennium} ({Cat}.
  {No}.{01CH37180})}}, Vol.~\bibinfo{volume}{4}. \bibinfo{pages}{2003--2011
  vol.4}.
\newblock
\urldef\tempurl%
\url{https://doi.org/10.1109/IROS.2001.976367}
\showDOI{\tempurl}


\bibitem[Mekonnen et~al\mbox{.}(2010)]%
        {mekonnen_constrained_2010}
\bibfield{author}{\bibinfo{person}{Zemene~W. Mekonnen}, \bibinfo{person}{Eric
  Slottke}, \bibinfo{person}{Heinrich Luecken}, \bibinfo{person}{Christoph
  Steiner}, {and} \bibinfo{person}{Armin Wittneben}.}
  \bibinfo{year}{2010}\natexlab{}.
\newblock \showarticletitle{Constrained maximum likelihood positioning for
  {UWB} based human motion tracking}. In \bibinfo{booktitle}{\emph{2010
  {International} {Conference} on {Indoor} {Positioning} and {Indoor}
  {Navigation}}}. \bibinfo{pages}{1--10}.
\newblock
\urldef\tempurl%
\url{https://doi.org/10.1109/IPIN.2010.5647912}
\showDOI{\tempurl}


\bibitem[Meyer et~al\mbox{.}(2015)]%
        {meyer_making_2015}
\bibfield{author}{\bibinfo{person}{Jochen Meyer}, \bibinfo{person}{Jutta
  Fortmann}, \bibinfo{person}{Merlin Wasmann}, {and} \bibinfo{person}{Wilko
  Heuten}.} \bibinfo{year}{2015}\natexlab{}.
\newblock \showarticletitle{Making {Lifelogging} {Usable}: {Design}
  {Guidelines} for {Activity} {Trackers}}. In
  \bibinfo{booktitle}{\emph{{MultiMedia} {Modeling}}}
  \emph{(\bibinfo{series}{Lecture {Notes} in {Computer} {Science}})},
  \bibfield{editor}{\bibinfo{person}{Xiangjian He}, \bibinfo{person}{Suhuai
  Luo}, \bibinfo{person}{Dacheng Tao}, \bibinfo{person}{Changsheng Xu},
  \bibinfo{person}{Jie Yang}, {and} \bibinfo{person}{Muhammad~Abul Hasan}}
  (Eds.). \bibinfo{publisher}{Springer International Publishing},
  \bibinfo{address}{Cham}, \bibinfo{pages}{323--334}.
\newblock
\showISBNx{978-3-319-14442-9}
\urldef\tempurl%
\url{https://doi.org/10.1007/978-3-319-14442-9_39}
\showDOI{\tempurl}


\bibitem[Miezal et~al\mbox{.}(2013)]%
        {miezal_generic_2013}
\bibfield{author}{\bibinfo{person}{Markus Miezal}, \bibinfo{person}{Gabriele
  Bleser}, \bibinfo{person}{Norbert Schmitz}, {and} \bibinfo{person}{Didier
  Stricker}.} \bibinfo{year}{2013}\natexlab{}.
\newblock \showarticletitle{A generic approach to inertial tracking of
  arbitrary kinematic chains}. In \bibinfo{booktitle}{\emph{Proceedings of the
  8th {International} {Conference} on {Body} {Area} {Networks}}}
  \emph{(\bibinfo{series}{{BodyNets} '13})}. \bibinfo{publisher}{ICST
  (Institute for Computer Sciences, Social-Informatics and Telecommunications
  Engineering)}, \bibinfo{address}{Brussels, BEL}, \bibinfo{pages}{189--192}.
\newblock
\showISBNx{978-1-936968-89-3}
\urldef\tempurl%
\url{https://doi.org/10.4108/icst.bodynets.2013.253608}
\showDOI{\tempurl}


\bibitem[Mollyn et~al\mbox{.}(2023)]%
        {mollyn_imuposer_2023}
\bibfield{author}{\bibinfo{person}{Vimal Mollyn}, \bibinfo{person}{Riku
  Arakawa}, \bibinfo{person}{Mayank Goel}, \bibinfo{person}{Chris Harrison},
  {and} \bibinfo{person}{Karan Ahuja}.} \bibinfo{year}{2023}\natexlab{}.
\newblock \showarticletitle{IMUPoser: Full-Body Pose Estimation using IMUs in
  Phones, Watches, and Earbuds}. In \bibinfo{booktitle}{\emph{Proceedings of
  the 2023 {CHI} {Conference} on {Human} {Factors} in {Computing} {Systems}}}
  \emph{(\bibinfo{series}{{CHI} '23})}. \bibinfo{publisher}{Association for
  Computing Machinery}, \bibinfo{address}{New York, NY, USA},
  \bibinfo{pages}{1--12}.
\newblock
\showISBNx{978-1-4503-8096-6}
\urldef\tempurl%
\url{https://doi.org/10.1145/3544548.3581392}
\showDOI{\tempurl}


\bibitem[Nguyen et~al\mbox{.}(2011)]%
        {nguyen_wearable_2011}
\bibfield{author}{\bibinfo{person}{Kim~Doang Nguyen}, \bibinfo{person}{I-Ming
  Chen}, \bibinfo{person}{Zhiqiang Luo}, \bibinfo{person}{Song~Huat Yeo}, {and}
  \bibinfo{person}{Henry Been-Lirn Duh}.} \bibinfo{year}{2011}\natexlab{}.
\newblock \showarticletitle{A {Wearable} {Sensing} {System} for {Tracking} and
  {Monitoring} of {Functional} {Arm} {Movement}}.
\newblock \bibinfo{journal}{\emph{IEEE/ASME Transactions on Mechatronics}}
  \bibinfo{volume}{16}, \bibinfo{number}{2} (\bibinfo{date}{April}
  \bibinfo{year}{2011}), \bibinfo{pages}{213--220}.
\newblock
\showISSN{1941-014X}
\urldef\tempurl%
\url{https://doi.org/10.1109/TMECH.2009.2039222}
\showDOI{\tempurl}
\newblock
\shownote{Conference Name: IEEE/ASME Transactions on Mechatronics}.


\bibitem[Rajanna et~al\mbox{.}(2014)]%
        {rajanna_step_2014}
\bibfield{author}{\bibinfo{person}{Vijay Rajanna}, \bibinfo{person}{Raniero
  Lara-Garduno}, \bibinfo{person}{Dev~Jyoti Behera}, \bibinfo{person}{Karthic
  Madanagopal}, \bibinfo{person}{Daniel Goldberg}, {and} \bibinfo{person}{Tracy
  Hammond}.} \bibinfo{year}{2014}\natexlab{}.
\newblock \showarticletitle{Step up life: a context aware health assistant}. In
  \bibinfo{booktitle}{\emph{Proceedings of the {Third} {ACM} {SIGSPATIAL}
  {International} {Workshop} on the {Use} of {GIS} in {Public} {Health}}}
  \emph{(\bibinfo{series}{{HealthGIS} '14})}. \bibinfo{publisher}{Association
  for Computing Machinery}, \bibinfo{address}{New York, NY, USA},
  \bibinfo{pages}{21--30}.
\newblock
\showISBNx{978-1-4503-3136-4}
\urldef\tempurl%
\url{https://doi.org/10.1145/2676629.2676636}
\showDOI{\tempurl}


\bibitem[Redmayne(2017)]%
        {redmayne_wheres_2017}
\bibfield{author}{\bibinfo{person}{Mary Redmayne}.}
  \bibinfo{year}{2017}\natexlab{}.
\newblock \showarticletitle{Where’s {Your} {Phone}? {A} {Survey} of {Where}
  {Women} {Aged} 15-40 {Carry} {Their} {Smartphone} and {Related} {Risk}
  {Perception}: {A} {Survey} and {Pilot} {Study}}.
\newblock \bibinfo{journal}{\emph{PLoS ONE}} \bibinfo{volume}{12},
  \bibinfo{number}{1} (\bibinfo{date}{Jan.} \bibinfo{year}{2017}),
  \bibinfo{pages}{e0167996}.
\newblock
\showISSN{1932-6203}
\urldef\tempurl%
\url{https://doi.org/10.1371/journal.pone.0167996}
\showDOI{\tempurl}


\bibitem[Ren et~al\mbox{.}(2022a)]%
        {ren_winect_2022}
\bibfield{author}{\bibinfo{person}{Yili Ren}, \bibinfo{person}{Zi Wang},
  \bibinfo{person}{Sheng Tan}, \bibinfo{person}{Yingying Chen}, {and}
  \bibinfo{person}{Jie Yang}.} \bibinfo{year}{2022}\natexlab{a}.
\newblock \showarticletitle{Winect: {3D} {Human} {Pose} {Tracking} for
  {Free}-form {Activity} {Using} {Commodity} {WiFi}}.
\newblock \bibinfo{journal}{\emph{Proceedings of the ACM on Interactive,
  Mobile, Wearable and Ubiquitous Technologies}} \bibinfo{volume}{5},
  \bibinfo{number}{4} (\bibinfo{date}{Dec.} \bibinfo{year}{2022}),
  \bibinfo{pages}{176:1--176:29}.
\newblock
\urldef\tempurl%
\url{https://doi.org/10.1145/3494973}
\showDOI{\tempurl}


\bibitem[Ren et~al\mbox{.}(2022b)]%
        {ren_gopose_2022}
\bibfield{author}{\bibinfo{person}{Yili Ren}, \bibinfo{person}{Zi Wang},
  \bibinfo{person}{Yichao Wang}, \bibinfo{person}{Sheng Tan},
  \bibinfo{person}{Yingying Chen}, {and} \bibinfo{person}{Jie Yang}.}
  \bibinfo{year}{2022}\natexlab{b}.
\newblock \showarticletitle{{GoPose}: {3D} {Human} {Pose} {Estimation} {Using}
  {WiFi}}.
\newblock \bibinfo{journal}{\emph{Proceedings of the ACM on Interactive,
  Mobile, Wearable and Ubiquitous Technologies}} \bibinfo{volume}{6},
  \bibinfo{number}{2} (\bibinfo{date}{July} \bibinfo{year}{2022}),
  \bibinfo{pages}{69:1--69:25}.
\newblock
\urldef\tempurl%
\url{https://doi.org/10.1145/3534605}
\showDOI{\tempurl}


\bibitem[Roetenberg et~al\mbox{.}(2013)]%
        {roetenberg_xsens_2013}
\bibfield{author}{\bibinfo{person}{Daniel Roetenberg}, \bibinfo{person}{Henk
  Luinge}, {and} \bibinfo{person}{Per Slycke}.}
  \bibinfo{year}{2013}\natexlab{}.
\newblock \showarticletitle{Xsens {MVN}: {Full} {6DOF} {Human} {Motion}
  {Tracking} {Using} {Miniature} {Inertial} {Sensors}}.
\newblock  (\bibinfo{year}{2013}), \bibinfo{pages}{9}.
\newblock


\bibitem[Roetenberg et~al\mbox{.}(2007)]%
        {roetenberg_ambulatory_2007}
\bibfield{author}{\bibinfo{person}{Daniel Roetenberg}, \bibinfo{person}{Per~J.
  Slycke}, {and} \bibinfo{person}{Peter~H. Veltink}.}
  \bibinfo{year}{2007}\natexlab{}.
\newblock \showarticletitle{Ambulatory {Position} and {Orientation} {Tracking}
  {Fusing} {Magnetic} and {Inertial} {Sensing}}.
\newblock \bibinfo{journal}{\emph{IEEE Transactions on Biomedical Engineering}}
  \bibinfo{volume}{54}, \bibinfo{number}{5} (\bibinfo{date}{May}
  \bibinfo{year}{2007}), \bibinfo{pages}{883--890}.
\newblock
\showISSN{1558-2531}
\urldef\tempurl%
\url{https://doi.org/10.1109/TBME.2006.889184}
\showDOI{\tempurl}
\newblock
\shownote{Conference Name: IEEE Transactions on Biomedical Engineering}.


\bibitem[Sengupta et~al\mbox{.}(2020)]%
        {sengupta_mm-pose_2020}
\bibfield{author}{\bibinfo{person}{Arindam Sengupta}, \bibinfo{person}{Feng
  Jin}, \bibinfo{person}{Renyuan Zhang}, {and} \bibinfo{person}{Siyang Cao}.}
  \bibinfo{year}{2020}\natexlab{}.
\newblock \showarticletitle{mm-{Pose}: {Real}-{Time} {Human} {Skeletal}
  {Posture} {Estimation} {Using} {mmWave} {Radars} and {CNNs}}.
\newblock \bibinfo{journal}{\emph{IEEE Sensors Journal}} \bibinfo{volume}{20},
  \bibinfo{number}{17} (\bibinfo{date}{Sept.} \bibinfo{year}{2020}),
  \bibinfo{pages}{10032--10044}.
\newblock
\showISSN{1558-1748}
\urldef\tempurl%
\url{https://doi.org/10.1109/JSEN.2020.2991741}
\showDOI{\tempurl}
\newblock
\shownote{Conference Name: IEEE Sensors Journal}.


\bibitem[Shen et~al\mbox{.}(2018)]%
        {shen_muse}
\bibfield{author}{\bibinfo{person}{Sheng Shen}, \bibinfo{person}{Mahanth
  Gowda}, {and} \bibinfo{person}{Romit Roy~Choudhury}.}
  \bibinfo{year}{2018}\natexlab{}.
\newblock \showarticletitle{Closing the Gaps in Inertial Motion Tracking}. In
  \bibinfo{booktitle}{\emph{Proceedings of the 24th Annual International
  Conference on Mobile Computing and Networking}} (New Delhi, India)
  \emph{(\bibinfo{series}{MobiCom '18})}. \bibinfo{publisher}{Association for
  Computing Machinery}, \bibinfo{address}{New York, NY, USA},
  \bibinfo{pages}{429–444}.
\newblock
\showISBNx{9781450359030}
\urldef\tempurl%
\url{https://doi.org/10.1145/3241539.3241582}
\showDOI{\tempurl}


\bibitem[Shen et~al\mbox{.}(2016)]%
        {shen_i_2016}
\bibfield{author}{\bibinfo{person}{Sheng Shen}, \bibinfo{person}{He Wang},
  {and} \bibinfo{person}{Romit Roy~Choudhury}.}
  \bibinfo{year}{2016}\natexlab{}.
\newblock \showarticletitle{I am a {Smartwatch} and {I} can {Track} my {User}'s
  {Arm}}. In \bibinfo{booktitle}{\emph{Proceedings of the 14th {Annual}
  {International} {Conference} on {Mobile} {Systems}, {Applications}, and
  {Services}}} \emph{(\bibinfo{series}{{MobiSys} '16})}.
  \bibinfo{publisher}{Association for Computing Machinery},
  \bibinfo{address}{New York, NY, USA}, \bibinfo{pages}{85--96}.
\newblock
\showISBNx{978-1-4503-4269-8}
\urldef\tempurl%
\url{https://doi.org/10.1145/2906388.2906407}
\showDOI{\tempurl}


\bibitem[Shen and House(2017)]%
        {shen_hand-arm_2017}
\bibfield{author}{\bibinfo{person}{Shixin~(Cindy) Shen} {and}
  \bibinfo{person}{Ronald~A. House}.} \bibinfo{year}{2017}\natexlab{}.
\newblock \showarticletitle{Hand-arm vibration syndrome}.
\newblock \bibinfo{journal}{\emph{Canadian Family Physician}}
  \bibinfo{volume}{63}, \bibinfo{number}{3} (\bibinfo{date}{March}
  \bibinfo{year}{2017}), \bibinfo{pages}{206--210}.
\newblock
\showISSN{0008-350X}
\urldef\tempurl%
\url{https://www.ncbi.nlm.nih.gov/pmc/articles/PMC5349719/}
\showURL{%
\tempurl}


\bibitem[Song et~al\mbox{.}(2021)]%
        {song_through-wall_2021}
\bibfield{author}{\bibinfo{person}{Yongkun Song}, \bibinfo{person}{Tian Jin},
  \bibinfo{person}{Yongpeng Dai}, \bibinfo{person}{Yongping Song}, {and}
  \bibinfo{person}{Xiaolong Zhou}.} \bibinfo{year}{2021}\natexlab{}.
\newblock \showarticletitle{Through-{Wall} {Human} {Pose} {Reconstruction} via
  {UWB} {MIMO} {Radar} and {3D} {CNN}}.
\newblock \bibinfo{journal}{\emph{Remote Sensing}} \bibinfo{volume}{13},
  \bibinfo{number}{2} (\bibinfo{date}{Jan.} \bibinfo{year}{2021}),
  \bibinfo{pages}{241}.
\newblock
\showISSN{2072-4292}
\urldef\tempurl%
\url{https://doi.org/10.3390/rs13020241}
\showDOI{\tempurl}


\bibitem[Spencer(2020)]%
        {spencer_pocket_2020}
\bibfield{author}{\bibinfo{person}{Elizabeth Spencer}.}
  \bibinfo{year}{2020}\natexlab{}.
\newblock \showarticletitle{The {Pocket}: {A} {Hidden} {History} of {Women}’s
  {Lives}, 1660-1900}.
\newblock \bibinfo{journal}{\emph{Cultural and Social History}}
  \bibinfo{volume}{17}, \bibinfo{number}{4} (\bibinfo{date}{Aug.}
  \bibinfo{year}{2020}), \bibinfo{pages}{570--572}.
\newblock
\showISSN{1478-0038}
\urldef\tempurl%
\url{https://doi.org/10.1080/14780038.2020.1810955}
\showDOI{\tempurl}
\newblock
\shownote{Publisher: Routledge \_eprint:
  https://doi.org/10.1080/14780038.2020.1810955}.


\bibitem[tesych(2020)]%
        {tesych_about_2020}
\bibfield{author}{\bibinfo{person}{tesych}.} \bibinfo{year}{2020}\natexlab{}.
\newblock \bibinfo{title}{About {Azure} {Kinect} {DK}}.
\newblock
\newblock
\urldef\tempurl%
\url{https://docs.microsoft.com/en-us/azure/kinect-dk/about-azure-kinect-dk}
\showURL{%
\tempurl}


\bibitem[Vlasic et~al\mbox{.}(2007)]%
        {vlasic_practical_2007}
\bibfield{author}{\bibinfo{person}{Daniel Vlasic}, \bibinfo{person}{Rolf
  Adelsberger}, \bibinfo{person}{Giovanni Vannucci}, \bibinfo{person}{John
  Barnwell}, \bibinfo{person}{Markus Gross}, \bibinfo{person}{Wojciech
  Matusik}, {and} \bibinfo{person}{Jovan Popović}.}
  \bibinfo{year}{2007}\natexlab{}.
\newblock \showarticletitle{Practical motion capture in everyday surroundings}.
\newblock \bibinfo{journal}{\emph{ACM Transactions on Graphics}}
  \bibinfo{volume}{26}, \bibinfo{number}{3} (\bibinfo{date}{July}
  \bibinfo{year}{2007}), \bibinfo{pages}{35}.
\newblock
\showISSN{0730-0301, 1557-7368}
\urldef\tempurl%
\url{https://doi.org/10.1145/1276377.1276421}
\showDOI{\tempurl}


\bibitem[von Marcard et~al\mbox{.}(2017)]%
        {von_marcard_sparse_2017}
\bibfield{author}{\bibinfo{person}{Timo von Marcard}, \bibinfo{person}{Bodo
  Rosenhahn}, \bibinfo{person}{Michael~J. Black}, {and} \bibinfo{person}{Gerard
  Pons-Moll}.} \bibinfo{year}{2017}\natexlab{}.
\newblock \bibinfo{title}{Sparse {Inertial} {Poser}: {Automatic} {3D} {Human}
  {Pose} {Estimation} from {Sparse} {IMUs}}.
\newblock
\newblock
\urldef\tempurl%
\url{http://arxiv.org/abs/1703.08014}
\showURL{%
\tempurl}
\newblock
\shownote{arXiv:1703.08014 [cs]}.


\bibitem[Wang et~al\mbox{.}(2018)]%
        {wang_rf-kinect_2018}
\bibfield{author}{\bibinfo{person}{Chuyu Wang}, \bibinfo{person}{Jian Liu},
  \bibinfo{person}{Yingying Chen}, \bibinfo{person}{Lei Xie},
  \bibinfo{person}{Hong~Bo Liu}, {and} \bibinfo{person}{Sanclu Lu}.}
  \bibinfo{year}{2018}\natexlab{}.
\newblock \showarticletitle{{RF}-{Kinect}: {A} {Wearable} {RFID}-based
  {Approach} {Towards} {3D} {Body} {Movement} {Tracking}}.
\newblock \bibinfo{journal}{\emph{Proceedings of the ACM on Interactive,
  Mobile, Wearable and Ubiquitous Technologies}} \bibinfo{volume}{2},
  \bibinfo{number}{1} (\bibinfo{date}{March} \bibinfo{year}{2018}),
  \bibinfo{pages}{1--28}.
\newblock
\showISSN{2474-9567}
\urldef\tempurl%
\url{https://doi.org/10.1145/3191773}
\showDOI{\tempurl}


\bibitem[Wei et~al\mbox{.}(2021)]%
        {wei_real_time}
\bibfield{author}{\bibinfo{person}{Wenchuan Wei}, \bibinfo{person}{Keiko
  Kurita}, \bibinfo{person}{Jilong Kuang}, {and} \bibinfo{person}{Alex Gao}.}
  \bibinfo{year}{2021}\natexlab{}.
\newblock \showarticletitle{Real-Time 3D Arm Motion Tracking Using the 6-axis
  IMU Sensor of a Smartwatch}. In \bibinfo{booktitle}{\emph{2021 IEEE 17th
  International Conference on Wearable and Implantable Body Sensor Networks
  (BSN)}}. \bibinfo{pages}{1--4}.
\newblock
\urldef\tempurl%
\url{https://doi.org/10.1109/BSN51625.2021.9507012}
\showDOI{\tempurl}


\bibitem[Wiese et~al\mbox{.}(2013)]%
        {wiese_phoneprioception_2013}
\bibfield{author}{\bibinfo{person}{Jason Wiese}, \bibinfo{person}{T.~Scott
  Saponas}, {and} \bibinfo{person}{A.J.~Bernheim Brush}.}
  \bibinfo{year}{2013}\natexlab{}.
\newblock \showarticletitle{Phoneprioception: enabling mobile phones to infer
  where they are kept}. In \bibinfo{booktitle}{\emph{Proceedings of the
  {SIGCHI} {Conference} on {Human} {Factors} in {Computing} {Systems}}}.
  \bibinfo{publisher}{ACM}, \bibinfo{address}{Paris France},
  \bibinfo{pages}{2157--2166}.
\newblock
\showISBNx{978-1-4503-1899-0}
\urldef\tempurl%
\url{https://doi.org/10.1145/2470654.2481296}
\showDOI{\tempurl}


\bibitem[Wikipedia(2022a)]%
        {noauthor_list_2022-1}
\bibfield{author}{\bibinfo{person}{Wikipedia}.}
  \bibinfo{year}{2022}\natexlab{a}.
\newblock \bibinfo{title}{List of best-selling video games}.
\newblock
\newblock
\urldef\tempurl%
\url{https://en.wikipedia.org/w/index.php?title=List_of_best-selling_video_games&oldid=1104319901}
\showURL{%
\tempurl}
\newblock
\shownote{Page Version ID: 1104319901}.


\bibitem[Wikipedia(2022b)]%
        {listofUWBWiki}
\bibfield{author}{\bibinfo{person}{Wikipedia}.}
  \bibinfo{year}{2022}\natexlab{b}.
\newblock \bibinfo{title}{List of {UWB}-enabled mobile devices}.
\newblock
\newblock
\urldef\tempurl%
\url{https://en.wikipedia.org/w/index.php?title=List_of_UWB-enabled_mobile_devices&oldid=1109918576}
\showURL{%
\tempurl}
\newblock
\shownote{Page Version ID: 1109918576}.


\bibitem[Wilk et~al\mbox{.}(2021)]%
        {wilk_multimodal_2021}
\bibfield{author}{\bibinfo{person}{Mariusz~P. Wilk}, \bibinfo{person}{Michael
  Walsh}, {and} \bibinfo{person}{Brendan O’Flynn}.}
  \bibinfo{year}{2021}\natexlab{}.
\newblock \showarticletitle{Multimodal {Sensor} {Fusion} for {Low}-{Power}
  {Wearable} {Human} {Motion} {Tracking} {Systems} in {Sports} {Applications}}.
\newblock \bibinfo{journal}{\emph{IEEE Sensors Journal}} \bibinfo{volume}{21},
  \bibinfo{number}{4} (\bibinfo{date}{Feb.} \bibinfo{year}{2021}),
  \bibinfo{pages}{5195--5212}.
\newblock
\showISSN{1558-1748}
\urldef\tempurl%
\url{https://doi.org/10.1109/JSEN.2020.3030779}
\showDOI{\tempurl}
\newblock
\shownote{Conference Name: IEEE Sensors Journal}.


\bibitem[Wittmann et~al\mbox{.}(2019)]%
        {wittmann_magnetometer-based_2019}
\bibfield{author}{\bibinfo{person}{Frieder Wittmann}, \bibinfo{person}{Olivier
  Lambercy}, {and} \bibinfo{person}{Roger Gassert}.}
  \bibinfo{year}{2019}\natexlab{}.
\newblock \showarticletitle{Magnetometer-{Based} {Drift} {Correction} {During}
  {Rest} in {IMU} {Arm} {Motion} {Tracking}}.
\newblock \bibinfo{journal}{\emph{Sensors}} \bibinfo{volume}{19},
  \bibinfo{number}{6} (\bibinfo{date}{Jan.} \bibinfo{year}{2019}),
  \bibinfo{pages}{1312}.
\newblock
\showISSN{1424-8220}
\urldef\tempurl%
\url{https://doi.org/10.3390/s19061312}
\showDOI{\tempurl}


\bibitem[Xiao and Zarar(2018)]%
        {xiao_wearable_2018}
\bibfield{author}{\bibinfo{person}{Xuesu Xiao} {and} \bibinfo{person}{Shuayb
  Zarar}.} \bibinfo{year}{2018}\natexlab{}.
\newblock \showarticletitle{A {Wearable} {System} for {Articulated} {Human}
  {Pose} {Tracking} {Under} {Uncertainty} of {Sensor} {Placement}}. In
  \bibinfo{booktitle}{\emph{2018 7th {IEEE} {International} {Conference} on
  {Biomedical} {Robotics} and {Biomechatronics} ({Biorob})}}.
  \bibinfo{pages}{1144--1150}.
\newblock
\urldef\tempurl%
\url{https://doi.org/10.1109/BIOROB.2018.8487858}
\showDOI{\tempurl}
\newblock
\shownote{ISSN: 2155-1782}.


\bibitem[Yang et~al\mbox{.}(2022)]%
        {yang_environment_2022}
\bibfield{author}{\bibinfo{person}{Chao Yang}, \bibinfo{person}{Lingxiao Wang},
  \bibinfo{person}{Xuyu Wang}, {and} \bibinfo{person}{Shiwen Mao}.}
  \bibinfo{year}{2022}\natexlab{}.
\newblock \showarticletitle{Environment {Adaptive} {RFID}-{Based} {3D} {Human}
  {Pose} {Tracking} {With} a {Meta}-{Learning} {Approach}}.
\newblock \bibinfo{journal}{\emph{IEEE Journal of Radio Frequency
  Identification}}  \bibinfo{volume}{6} (\bibinfo{year}{2022}),
  \bibinfo{pages}{413--425}.
\newblock
\showISSN{2469-7281}
\urldef\tempurl%
\url{https://doi.org/10.1109/JRFID.2022.3140256}
\showDOI{\tempurl}
\newblock
\shownote{Conference Name: IEEE Journal of Radio Frequency Identification}.


\bibitem[Yi et~al\mbox{.}(2022)]%
        {yi_physical_2022}
\bibfield{author}{\bibinfo{person}{Xinyu Yi}, \bibinfo{person}{Yuxiao Zhou},
  \bibinfo{person}{Vladislav Golyanik}, \bibinfo{person}{Marc Habermann},
  \bibinfo{person}{Soshi Shimada}, \bibinfo{person}{Christian Theobalt}, {and}
  \bibinfo{person}{Feng Xu}.} \bibinfo{year}{2022}\natexlab{}.
\newblock \showarticletitle{Physical {Inertial} {Poser} ({PIP}):
  {Physics}-aware {Real}-time {Human} {Motion} {Tracking} from {Sparse}
  {Inertial} {Sensors}}.
\newblock  (\bibinfo{date}{June} \bibinfo{year}{2022}), \bibinfo{pages}{15}.
\newblock


\bibitem[Yi et~al\mbox{.}(2021)]%
        {yi_transpose_2021}
\bibfield{author}{\bibinfo{person}{Xinyu Yi}, \bibinfo{person}{Yuxiao Zhou},
  {and} \bibinfo{person}{Feng Xu}.} \bibinfo{year}{2021}\natexlab{}.
\newblock \showarticletitle{{TransPose}: real-time {3D} human translation and
  pose estimation with six inertial sensors}.
\newblock \bibinfo{journal}{\emph{ACM Transactions on Graphics}}
  \bibinfo{volume}{40}, \bibinfo{number}{4} (\bibinfo{date}{July}
  \bibinfo{year}{2021}), \bibinfo{pages}{86:1--86:13}.
\newblock
\showISSN{0730-0301}
\urldef\tempurl%
\url{https://doi.org/10.1145/3450626.3459786}
\showDOI{\tempurl}


\bibitem[Young(2010)]%
        {young_use_2010}
\bibfield{author}{\bibinfo{person}{Alexander~David Young}.}
  \bibinfo{year}{2010}\natexlab{}.
\newblock \showarticletitle{Use of {Body} {Model} {Constraints} to {Improve}
  {Accuracy} of {Inertial} {Motion} {Capture}}. In
  \bibinfo{booktitle}{\emph{2010 {International} {Conference} on {Body}
  {Sensor} {Networks}}}. \bibinfo{pages}{180--186}.
\newblock
\urldef\tempurl%
\url{https://doi.org/10.1109/BSN.2010.30}
\showDOI{\tempurl}
\newblock
\shownote{ISSN: 2376-8894}.


\bibitem[Zhou et~al\mbox{.}(2020)]%
        {zhou_limbmotion}
\bibfield{author}{\bibinfo{person}{Han Zhou}, \bibinfo{person}{Yi Gao},
  \bibinfo{person}{Xinyi Song}, \bibinfo{person}{Wenxin Liu}, {and}
  \bibinfo{person}{Wei Dong}.} \bibinfo{year}{2020}\natexlab{}.
\newblock \showarticletitle{LimbMotion: Decimeter-Level Limb Tracking for
  Wearable-Based Human-Computer Interaction}.
\newblock \bibinfo{journal}{\emph{Proc. ACM Interact. Mob. Wearable Ubiquitous
  Technol.}} \bibinfo{volume}{3}, \bibinfo{number}{4}, Article
  \bibinfo{articleno}{161} (\bibinfo{date}{sep} \bibinfo{year}{2020}),
  \bibinfo{numpages}{24}~pages.
\newblock
\urldef\tempurl%
\url{https://doi.org/10.1145/3369836}
\showDOI{\tempurl}


\bibitem[Zhou et~al\mbox{.}(2008)]%
        {zhou_use_2008}
\bibfield{author}{\bibinfo{person}{Huiyu Zhou}, \bibinfo{person}{Thomas Stone},
  \bibinfo{person}{Huosheng Hu}, {and} \bibinfo{person}{Nigel Harris}.}
  \bibinfo{year}{2008}\natexlab{}.
\newblock \showarticletitle{Use of multiple wearable inertial sensors in upper
  limb motion tracking}.
\newblock \bibinfo{journal}{\emph{Medical Engineering \& Physics}}
  \bibinfo{volume}{30}, \bibinfo{number}{1} (\bibinfo{date}{Jan.}
  \bibinfo{year}{2008}), \bibinfo{pages}{123--133}.
\newblock
\showISSN{1350-4533}
\urldef\tempurl%
\url{https://doi.org/10.1016/j.medengphy.2006.11.010}
\showDOI{\tempurl}


\bibitem[Zihajehzadeh et~al\mbox{.}(2015)]%
        {zihajehzadeh_uwb-aided_2015}
\bibfield{author}{\bibinfo{person}{Shaghayegh Zihajehzadeh},
  \bibinfo{person}{Paul~K. Yoon}, \bibinfo{person}{Bong-Soo Kang}, {and}
  \bibinfo{person}{Edward~J. Park}.} \bibinfo{year}{2015}\natexlab{}.
\newblock \showarticletitle{{UWB}-{Aided} {Inertial} {Motion} {Capture} for
  {Lower} {Body} 3-{D} {Dynamic} {Activity} and {Trajectory} {Tracking}}.
\newblock \bibinfo{journal}{\emph{IEEE Transactions on Instrumentation and
  Measurement}} \bibinfo{volume}{64}, \bibinfo{number}{12}
  (\bibinfo{date}{Dec.} \bibinfo{year}{2015}), \bibinfo{pages}{3577--3587}.
\newblock
\showISSN{1557-9662}
\urldef\tempurl%
\url{https://doi.org/10.1109/TIM.2015.2459532}
\showDOI{\tempurl}


\end{thebibliography}

\end{document}